\documentclass[twocolumn,preprint]{aastex6}

\slugcomment{Original paper}

\shorttitle{The source structure of 0642+449}
\shortauthors{Xu et al.}

\begin{document}


\title{The source structure of 0642+449 detected from the CONT14 observations}


\author{Ming H. Xu\altaffilmark{1,2}, Robert Heinkelmann\altaffilmark{2}, James M. Anderson\altaffilmark{2}, Julian Mora-Diaz\altaffilmark{2}, Harald Schuh\altaffilmark{2,3}, and Guang L. Wang\altaffilmark{1}}
%

%
\altaffiltext{1}{Shanghai Astronomical Observatory, Chinese Academy of Sciences,
               No. 80 Nandan Raod, 200030, Shanghai, China; mhxu@shao.ac.cn}
\altaffiltext{2}{DeutschesGeoForschungsZentrum (GFZ), Potsdam,
            Telegrafenberg, 14473 Potsdam, Germany}
\altaffiltext{3}{Institute of Geodesy and Geoinformation Science, 
      Technische Universit\"{a}t Berlin, Stra${\beta}$e des 17. Juni 135, 10623, Berlin, Germany}
%

\begin{abstract}
The CONT14 campaign with state-of-the-art VLBI data has
observed the source 0642+449 with about one thousand observables
each day during a continuous observing period of fifteen days,
providing tens of thousands of closure delays---the sum of the delays around a 
closed loop of baselines.
The closure delay is independent of the instrumental and propagation delays and provides valuable
additional information about the source structure. 
We demonstrate the use of this new ``observable'' for the determination of the structure 
in the radio source 0642+449.
This source, as one of the defining sources in the second 
realization of the International Celestial Reference Frame (ICRF2), 
is found to have two point-like components with
a relative position offset of $-$426 microarcseconds ($\mu$as) in right ascension and $-$66 $\mu$as
in declination. The two components are almost equally bright with a
flux-density ratio of 0.92. 
The standard deviation of closure delays for source 0642+449
was reduced from 139 ps to 90 ps by using this two-component model. Closure
delays larger than one nanosecond are found to be related to the source
structure, demonstrating that structure effects for a source
with this simple structure could be up to tens of nanoseconds. The method described in this paper 
does not rely on a priori source structure information, such as knowledge of
source structure determined from direct (Fourier) imaging of the same observations
or observations at other epochs. We anticipate our study to be a starting point
for more effective determination of the structure effect in VLBI observations.
\end{abstract}





\section{Introduction}

Radio galaxies and quasars have radio-emitting structure that can be conveniently divided into two
categories: extended structure, the dimensions of which range
from $10^{3}$ pc to even $10^{6}$ pc, and compact structure, 
with dimensions typically ranging from 1 pc to 100 pc \citep{kel88}.
Extragalactic radio sources with compact structure are used to
realize the fundamental Celestial Reference Frame with axis
stability at the level of ten microarcseconds ($\mu$as) by very
long baseline interferometry (VLBI) observations \citep{ma98, fey15}.
Given that the typical distance to these sources is at the level of $10^{9}$~pc, the
compact structure should have angular dimensions of
0.2--20 milliarcseconds (mas), as shown in images of
astrometric sources from astrophysical imaging studies
\citep[e.g.,][]{cha90a, ojh04, ojh05, pin07, lis09, cha10, lis13}. For example,
survey images of 91 compact sources obtained from VLBI observations at 5 GHz
by \citet{tay94} showed that only eight sources had a structure
smaller than one milliarcsecond. The effects of source structure on source
position determined from VLBI observations were studied and
demonstrated in a series of studies \citep[e.g.,][]{whi71, fey97, fey00,
fei03, mac07, mal08, moo11}. Recently, by observing four close radio
sources in the second 
realization of the International Celestial Reference Frame (ICRF2) 
for five times over one year, \citet{fom11} found that the
radio flux intensity maximum could follow a jet component rather
than stay close to the radio core. Their study suggests that if
the jet component gets fainter than the radio core or if they get completely
separated at some time, significant position variations will occur
at the level of 0.1 mas yr$^{-1}$ or even larger.

The study of the source structure effect on VLBI observables was
pioneered by \citet{tho80}. A significant
effort was made by \citet{cha90b}, who modeled the source structure
corrections for VLBI group delay and phase rate observables based on
the brightness distributions of the sources. Many studies then 
attempted to introduce the theoretical model of the structure effect 
into astrometric VLBI data analysis based on images of sources
\citep[e.g.,][]{cam88, cha88, tan88, ulv88, zep91, cha93, gon93, fey96, pet07}. 
For example, \citet{sov02} applied it to a series of
ten of the Research and Development VLBI (RDV) sessions, 
and the results showed that the weighted delay residuals could be reduced. 
An example of the application of the theoretical
model to the European geodetic VLBI sessions was tested by
\citet{tor07}.

There are, however, several points that presently limit the
application of this model for the correction of the structure effect. First, the source
structure effect is very sensitive to a slight change in the brightness
distribution. Unfortunately, the time histories of available images for most
sources are quite sparse, and in the foreseeable future it is almost impossible
to make images on regular basis at intervals of much less than a year 
for so many sources in the astrometric catalog unless astrometric/geodetic observations
themselves will be scheduled in a suitable way and sufficient efforts of making images
will be made. Secondly, even when images made several
months apart from each other are available, the stationary
reference point in these images can be hard to recognize
if the radio flux intensity maximum observed by VLBI is dominated
by a jet component. Consequently, in standard astrometric/geodetic VLBI data
analysis, the source structure effect has not actually been handled
so far. The source structure effect is still very important and
challenging for the astrometric VLBI, as shown in
simulation studies \citep{sha15, pla16}. If VLBI is to achieve its full potential
of the realization of the extragalactic Celestial Reference Frame with accuracy of the microarcsecond 
level and that of the Terrestrial Reference Frame with accuracy of the millimeter level, 
it is necessary to study and handle the source structure effect more effectively based on
the astrometric observations themselves.
These are the purposes of this paper.

In this paper we perform an initial analysis to determine how well
source structure can be determined directly from the geodetic VLBI
observables themselves \footnote{\added{Geodetic/astrometric VLBI observables are the baseline-based group delays and phase rates determined per scan within a geodetic VLBI experiment. The International VLBI Service for Geodesy and Astrometry (IVS) coordinates archives of geodetic VLBI experiment observables (see http://ivscc.gsfc.nasa.gov/products-data/data.html), but visibility datasets are normally not made available for analysis.}}. We aim to develop an alternative method for
studying the structure effect that should be simple, easy to
implement, and applicable for general historical and future geodetic
VLBI observations, including many of the oldest observations (back to
the 1970s) for which the visibility datasets are no longer available.
Although a self-calibration and Fourier imaging analysis of the
visibility data can give superior results for determining source
structure, that approach is time and computing resource intensive, it
requires large amounts of software not currently implemented in
geodetic analysis packages, it requires that the observations were
conducted in a manner suitable for imaging, which is frequently not
the case for historical geodetic VLBI sessions, and it will be
difficult, yield sub-optimal results, or even be impossible for the
historical experiments that no longer have archived visibility
datasets.

Therefore we defer our structure analysis based on imaging for a
future publication, and
we make use of the closure delay, the sum of the delays around a closed 
loop of baselines, as a new observable and propose 
a method to use this new observable for the determination of
the source structure effect on the astrometric VLBI observable. 
We calculate the closure delays, investigate the
characteristics of the source structure, and
then solve for the source structure effect on each observable. 
The source structure can be finally
obtained and the source structure effect can be determined. 
The source 0642+449, one of the ICRF2 defining
sources, is selected as a demonstration case for this method.

The systematical analysis of closure delay requires a consistent definition and 
a careful discussion of closure delay,
which are presented along with its calculating model 
in Section 2. The data used here, the CONT14 observations, and the overall
statistics of the closure delay of source 0642+449 are
introduced in Section 3. Section 4 discusses the method that was
used to solve for the source structure effect on each observable
based on the knowledge from Section 3. The results, describing 
the structure of this source, are shown in Section 5,
and the final model is presented in Section 6. Conclusions and discussion are given 
in the last Section.

\section{Model of closure delay}


The closure phase, 
unaffected by instrumental and atmospheric instabilities, 
was recognized as a good observable for the study of source structure
first by \citet{jen58} and later by \citet{rog74}. 
A method for
recovering the brightness distributions of compact radio sources 
from VLBI observations of closure phase, 
together with the measured visibility amplitudes, was developed to make images 
at the scale of milliarcseconds \citep{rea78} and 
used to obtain valuable maps for 45 objects by \citet{pea81, pea88}
in the early stage of VLBI imaging. The methods using the closure phase
for recovering the brightness distribution of a radio source were called \emph{hybrid-mapping based on closure quantities},
which was widely applied for the study of source structure \citep[see][and references therein]{pea84}. 
In terms of independence from instrumental and atmospheric instabilities and 
sensitivity to source structure,
closure delay has similar characteristics to closure phase. 
The closure delay was used for evaluating the performance of the structure models
in the past \citep{cha90b}, but our method will use it in a more direct way. 
For the first time, the definition and the model of closure delay is given here.

For a triangle of 3 stations, $a$, $b$, and $c$, the closure delay is defined for an individual
wavefront by
    \begin{equation}
     \label{eq_closure}
\tau_{abc}\equiv\tau_{ab}+\tau_{bc}+\tau_{ca},
    \end{equation}
where, for instance, $\tau_{ab}$ is the delay observable from station $a$ to
station $b$, and $\tau_{bc}$ is the delay observable from station $b$ to
station $c$, for the same wavefront received by three stations. 
Equation (\ref{eq_closure}) depicts the scenario in which
the same wavefront passes stations $a$, $b$, and $c$ in some arbitrary
sequence. The closure delay, for an ideal point source, is independent of any
station-based or source-based effect and should be zero if there is
no observational noise. For sources with detectable structure, as expected 
even for most astrometric sources, the group delay observable depends on the 
observing frequency and the baseline length and orientation differently
from the group delay behavior for a point source. In general, the change in
the group delay will be different for each baseline in the closure delay triangle
dependent upon the source structure, and the resulting closure delay will be non-zero.
Therefore, variations in the closure delays sufficiently far away from zero that they are
unlikely to be caused by random measurement noise should, in principle, only be observed
for sources with significant structure.

In geodetic VLBI measurements, by convention, the time tag of group delay, phase, and phase rate observables is
referred to the epoch when the wavefront passes the first station in the
baseline. They are implemented in the VLBI processor such that all observables in one scan have
exactly the same time tag value. This means that the observables in
one scan are usually not related to the same wavefront and allowance 
must be made for the delay of the arrival of the same wavefront at the different stations. 
Moreover, the baseline name is made up of two stations' names in the alphabetical order, 
so that $\tau_{ca}$ 
should be replaced with the actual geodetic VLBI observables on that baseline, $-\tau_{ac}$.
Therefore, to the accuracy of the
second order in delay, the closure delay at reference epoch $t$ is calculated from

    \begin{equation}
     \label{eq_closure2}
\tau_{abc}(t)=\tau_{ab}(t)+[\tau_{bc}(t)+\dot{\tau}_{bc}(t)\cdotp{\tau^{\prime}_{ab}(t)}+\frac{1}{2}\ddot{\tau}_{bc}(t)\cdotp{{\tau^{\prime}_{ab}(t)}^2}]-\tau_{ac}(t),
    \end{equation}
where a prime on a delay symbol indicates an absence of dependence
on station clock offset, that is, referring to the geometric delay, and
a superposed dot and double superposed dots denote differentiation
with respect to time once and twice, respectively.

For a goal of 1 ps accuracy of the closure delay, 
the third term inside the bracket of Equation (\ref{eq_closure2})
is smaller than 0.1 ps and is therefore negligible, 
since for ground-based VLBI the magnitude of the group delay is at the level of 0.02~s and that of the second order derivative is at the
level of 10~ps~s$^{-2}$. Since the
magnitude of the first order derivative is at the level of 1~$\mu$s~s$^{-1}$, the
second term inside the bracket has the magnitude of about ten nanoseconds (ns) and should be
calculated as accurately as possible. The $\dot{\tau}_{bc}(t)$ term can be
calculated from the theoretical model, which may have to ignore the
rate of change of the propagation delay; it can be derived more accurately from
the phase rate observable and the ionospheric delay rate,
both of which are available for International VLBI Service for
Geodesy and Astrometry \citep[IVS,][]{sch12} VLBI observations. All clock offsets or
jumps larger than 0.1 $\mu$s should be taken into account for the
geometric delay, $\tau^{\prime}_{ab}(t)$, in Equation (\ref{eq_closure2}) if it is calculated
from the group delay observable.

Following the conventions in geodetic VLBI measurements of the time tag and of
the baseline's name, the closure delay in this paper will be
referred to the same epoch as that of the three observables in the
triangle and be labeled by the names of the three stations in 
alphabetical order. In this case, the baseline between the
first station and the third station in a triangle will always
contribute to the closure delay calculated from Equation 
(\ref{eq_closure2}), in contrast to other two
baselines, with a negative sign.




\section{Closure delay of source 0642+449}

The data from CONT14\footnote{http://ivscc.gsfc.nasa.gov/program/cont14/}
observations \citep{not15} at X band were used.
CONT14 is a campaign of continuous VLBI observations, conducted by the IVS from
00:00:00 UT on 2014 May 6 to 00:00:00 UT on 2014 May 21. It is a
continuation of the series of continuous VLBI observations over 15 days that were
observed every three years since 2002. With a network of ten
stations in the northern hemisphere and seven in the southern
hemisphere at sixteen sites\footnote{Two stations, HOBART12 and HOBART26, 
are located at the Hobart site in Australia.}, this campaign was intended to
acquire state-of-the-art VLBI data over a time period of 15 days
with the highest accuracy that the then existing VLBI
system was capable. In these continuous observations there are
several sources with about one thousand observables per day, which makes
the CONT14 observations considerably valuable for the study of 
source structure.

\begin{figure}
\begin{center}

  \includegraphics[width=0.48\textwidth]{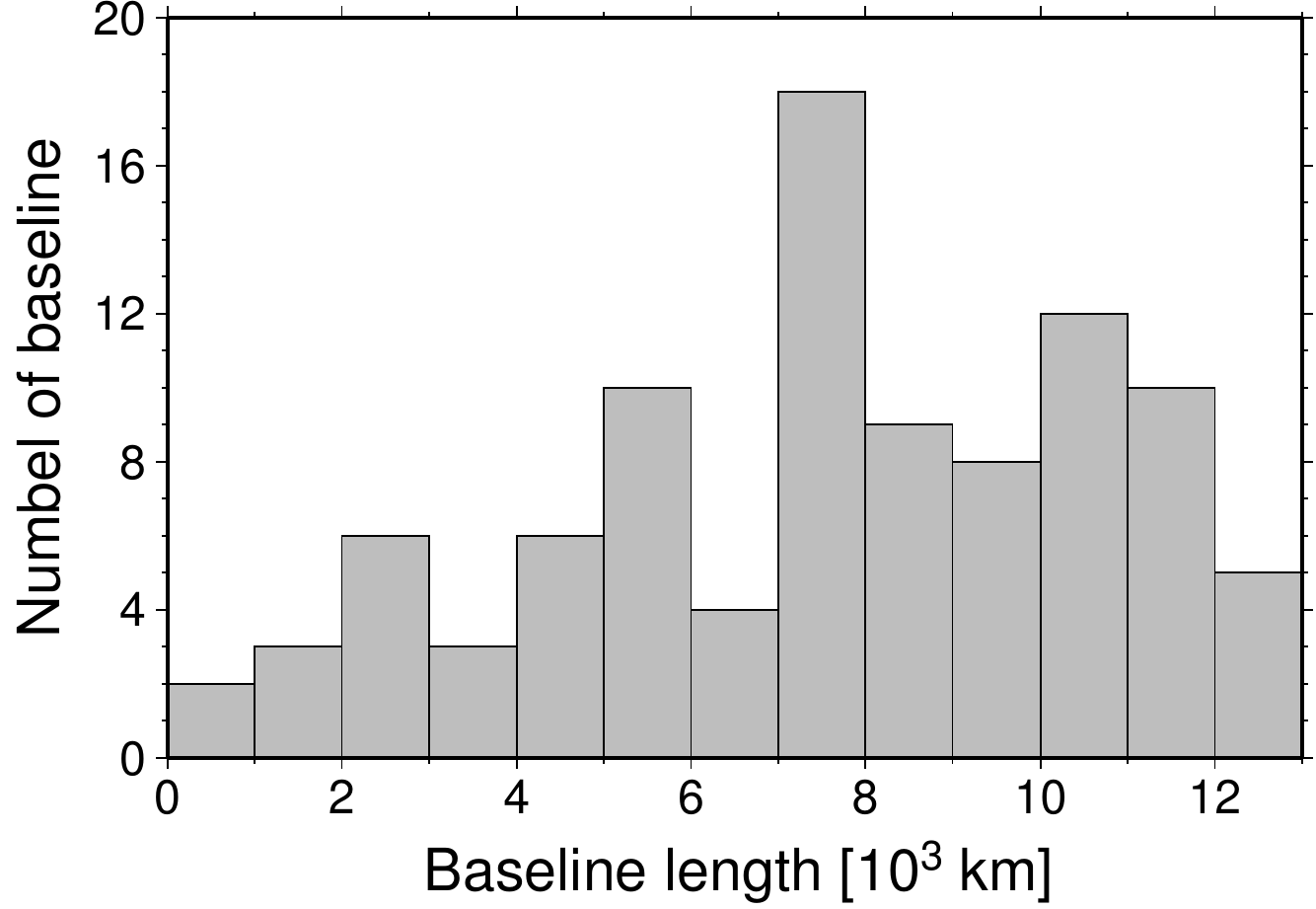}
  \caption{The baseline lengths of the observing network of source 0642+449 in CONT14. There were in total 15 stations
  observing this source, and 96 baselines were correlated.}
  \label{base_leng}
   
\end{center}

\end{figure}

The radio source 0642+449 was
selected as the target of our study. It was observed in 512 scans
by 15 stations in the CONT14 campaign (only the two southernmost
stations at the Hobart site could not observe this source), and has in total 11~027 pairs of
usable (quality codes\footnote{http://lupus.gsfc.nasa.gov/global/ngs-doc.html}
smaller than 8) group delay and delay rate
observables at X band. In total, 22~154 closure delays were
calculated from this data set and provided important statistics of
the performance of this individual source. As the number of triangles
is about two times that of group delay observables, each individual
group delay on average is involved in six triangles. The resulting
triangles have 350 different kinds of geometry. There are in total
96 baselines with lengths ranging from 900 km to 12~600~km as shown in
Figure \ref{base_leng}. This widely-spanned baseline length provides 
a good $uv$ coverage, eventually facilitating the detection of the source structure
at a variety of scales.

\begin{figure}
\begin{center}
   \includegraphics[width=0.48\textwidth]{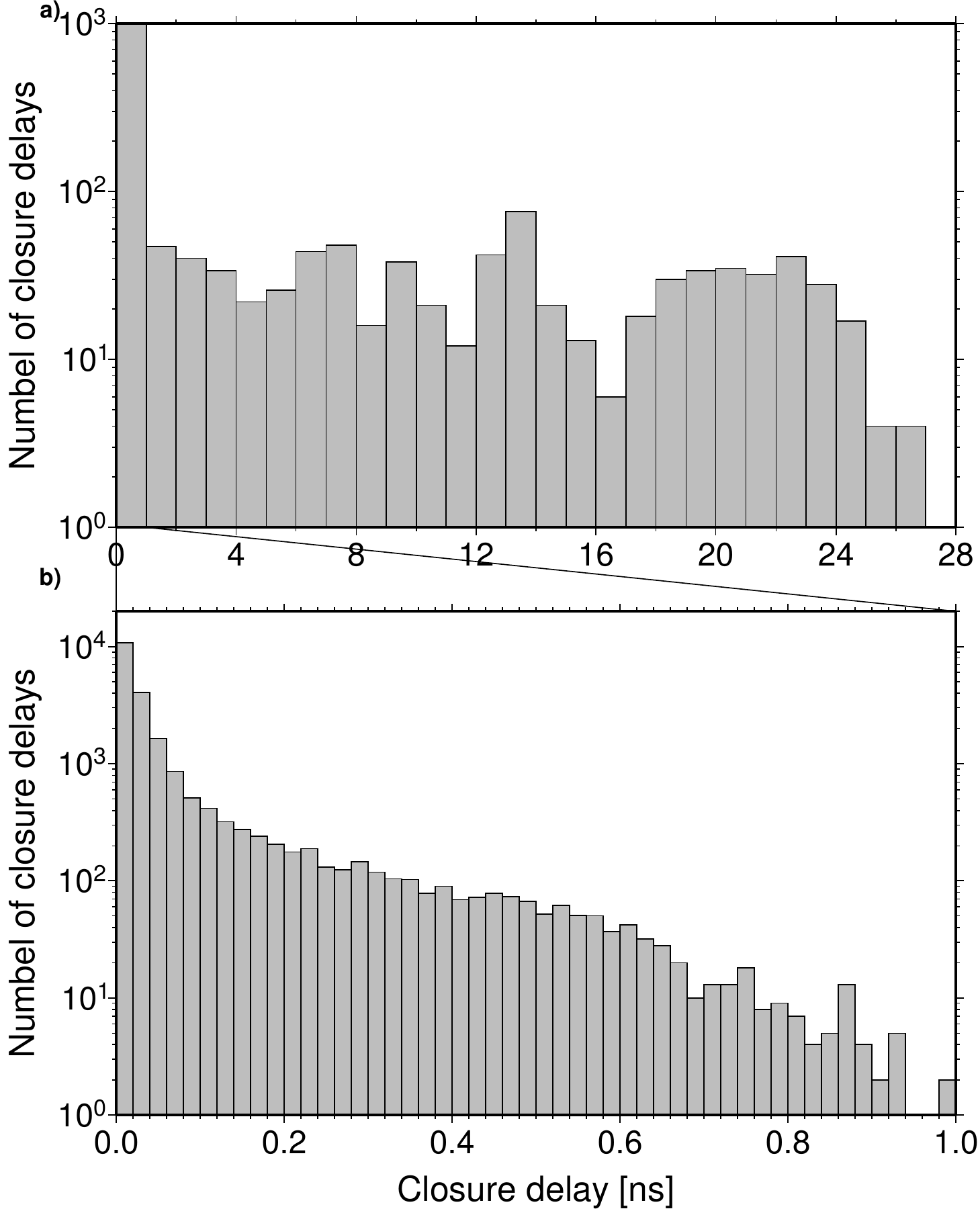}
  \caption{The absolute magnitude distribution of closure delay of the source 0642+449.
  The subplot $a$ shows the distribution of closure delay in
  the whole range from 0 ns to 28 ns with the bin width of 1 ns,
  and the subplot $b$ shows, with a smaller bin width of 20 ps to
  demonstrate more detail, the distribution in the main range from 0 ns to 1 ns
  which has been out of the axis limit as one bin in the subplot $b$.}
  \label{nemb0642}
   
\end{center}

\end{figure}

\begin{figure}
\begin{center}
 
  \includegraphics[width=0.47\textwidth]{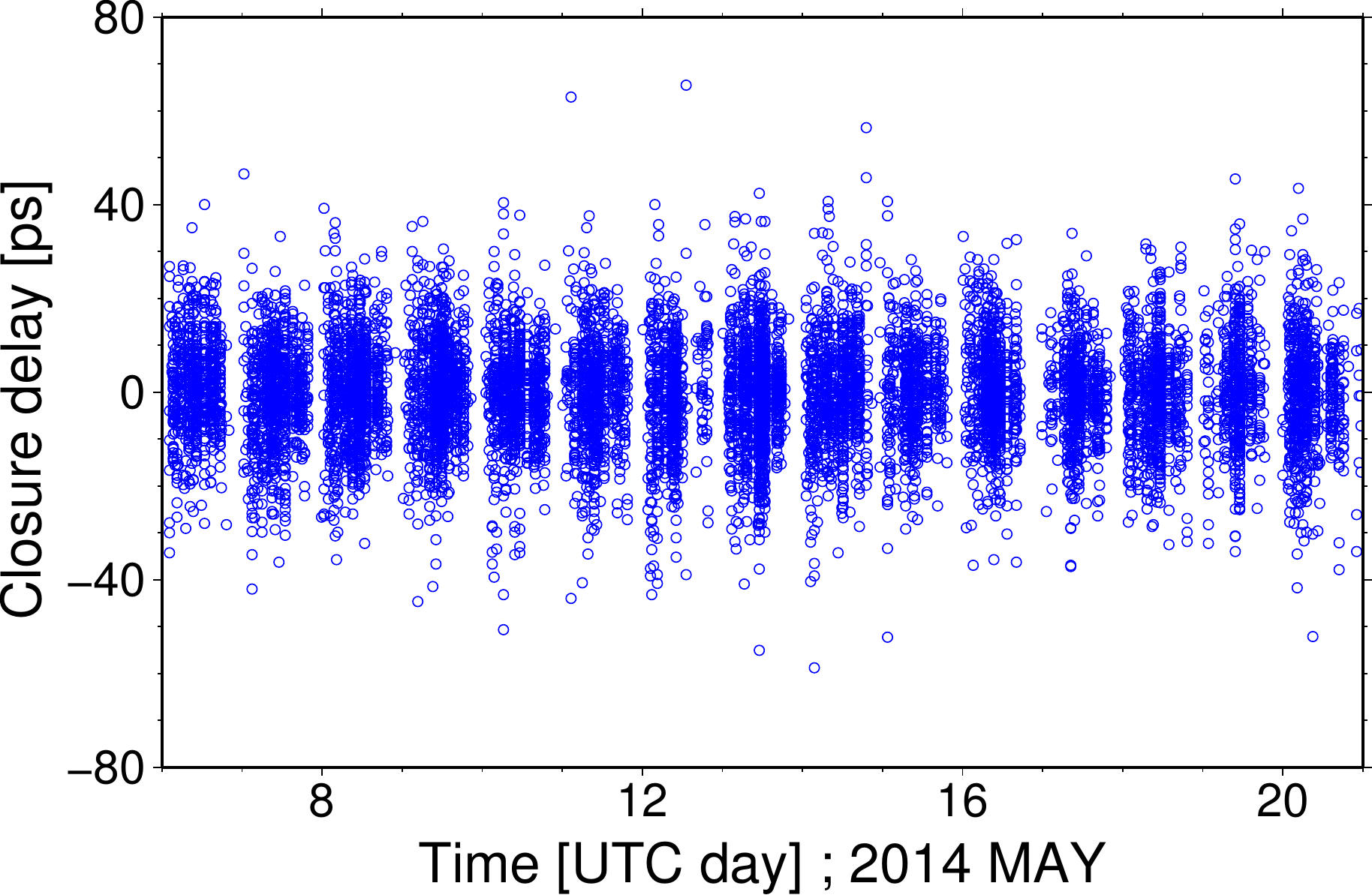}
  \caption{The closure delays of 0727-115 over 15 days. There is absolutely no triangle with closure delay of magnitude larger than 70 ps, and most triangles are with closure delays of magnitude smaller than 30 ps.}
  \label{s0727}

\end{center}

  \end{figure}

Multiband group delay ambiguities of 100 ns or 50 ns, determined by inspecting the closure
delay, were fixed for three group delays
for the baseline WESTFORD--YARRA12M on the 6th, 10th, and 11th of May. After this change to the CONT14 observations, the 22~154 closure
delays were then recalculated. Six closure delays
with absolute value larger than 28 ns were subsequently
excluded as outliers in the analysis here. No other changes to the data
were made and no additional points were excluded from analysis.
Figure \ref{nemb0642} shows the
distribution of the absolute magnitudes of the 22~148 closure delays. 
Approximately 49$\%$ of the triangles have closure
delays of absolute value smaller than 20 ps, and 67$\%$ of triangles have closure delays of that  
smaller than 40 ps. As Figure \ref{nemb0642} shows,
there is a rather flat distribution in each of the subplots: the one
in subplot $a$ lies in the range larger than
1 ns containing 759 closure delays; and the one in subplot $b$ lies 
in the range of 0.1 ns to 0.9 ns, which
has $16\%$ of triangles. The mean value and the
standard deviation of closure delays for the whole set of triangles is
$-$0.02 ns and 2.668 ns, respectively, and 0.3~ps and 139 ps, respectively,
for the set in the range 0 to 1.0 ns. The statistics and the
distribution of closure delay demonstrate that source 0642+449 to
some extent performs well for a geodetic source showing closure delays of less than 20 ps for half of the triangles,
but it is not as compact as a point-like source.
Furthermore, 
there are about 759 closure delay magnitudes above 1~ns, suggesting that about 
127 group delays (based on the average number of triangles each single 
group delay is part of), 
about 1~\% of the total observables for this source, sense exceptionally large structure effects.
For a comparison to demonstrate the measurement noise in VLBI observables, the 
standard deviations of closure delays for unresolved sources, such as 0016+731 and 0727-115, 
were calculated as well. Source 0016+731 has about 23~300 closure delays and 0727-115 has about
11~200 closure delays. The standard
deviation for source 0016+731, which showed
a little resolved structure, is about 11 ps, and that for source 0727-115 is about 8.0 ps. 
Figure \ref{s0727} shows the closure delay distribution over 15 days.
Geodetic VLBI observations are scheduled with
different observing durations for each station and each expected source brightness to achieve
a uniform signal to noise ratio for all observations, so we assume
that the 8.0/1.7=4.6~ps standard deviation for $0727-115$ represents the upper
limit to the typical measurement noise for all sources in CONT14 observations, and the vast
majority of the 139~ps scatter in the $0642+449$ closure delays cannot be 
explained by the measurement noise.

\subsection{Triangles with three short baselines}

\begin{figure}
\begin{center}
 
  \includegraphics[width=0.50\textwidth]{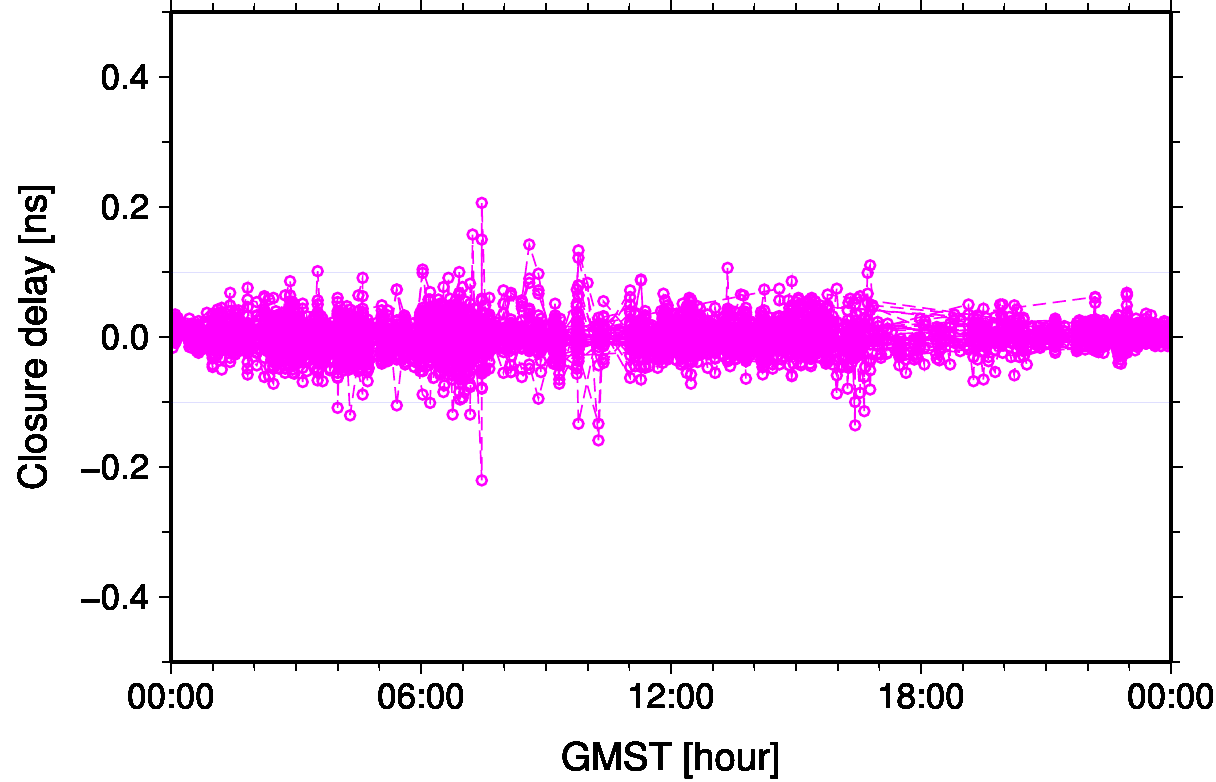}
  \caption{The closure delays of triangles with all three baselines
  shorter than 7100 km, which in total have 42 different geometries. All points are connected by dashed lines to
  show whether there are points out of the chosen axis limit. This is applied to all figures that show the closure delay or
  the source structure effect in the paper. For this case, not a single closure
  delay is out of the axis limit. }
  \label{small_net1}

\end{center}

  \end{figure}

\begin{figure}
\begin{center}
 
  \includegraphics[width=0.48\textwidth]{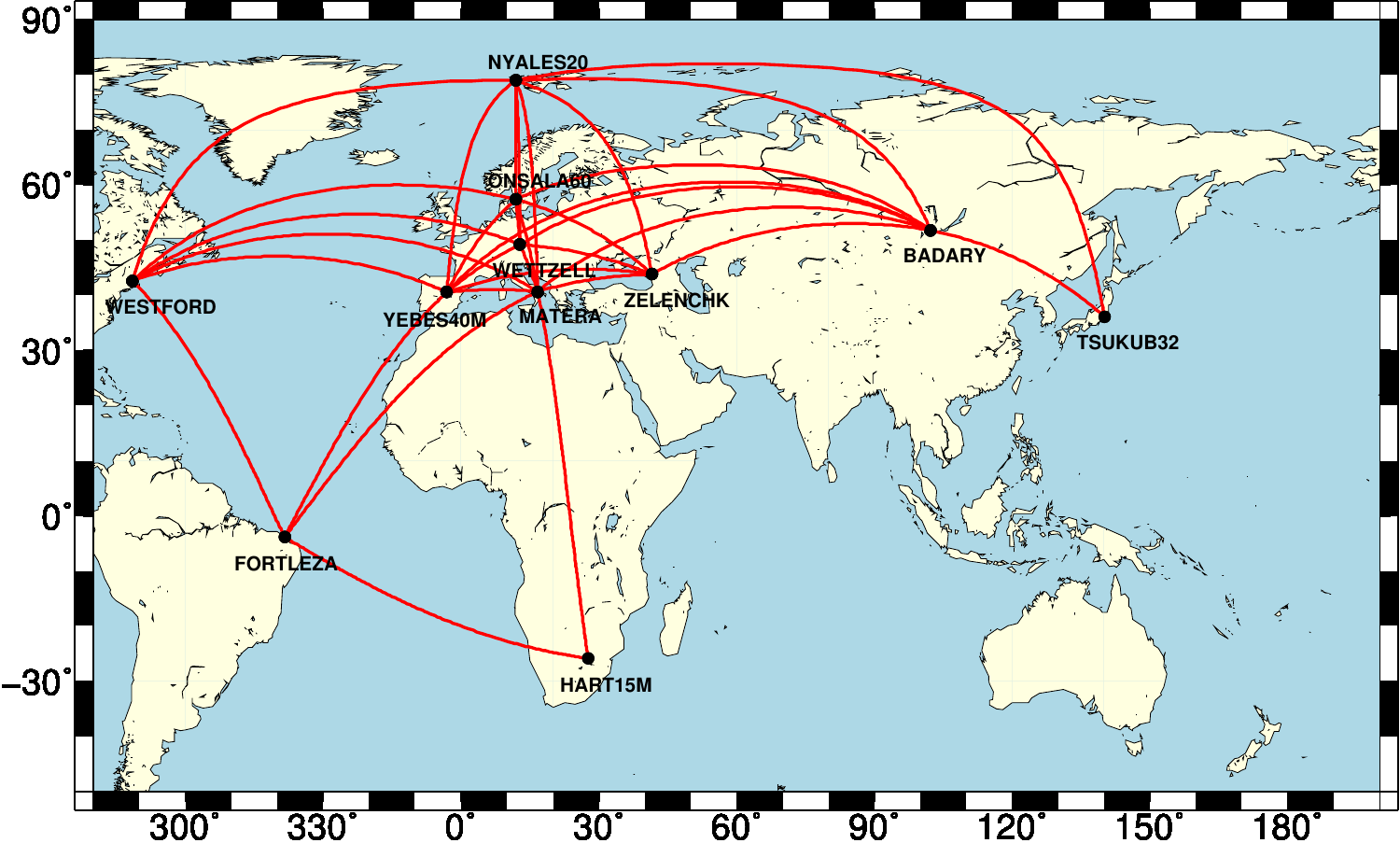}
  \caption{The triangles with all three baselines shorter than 7100 km.}
  \label{small_net2}

\end{center}

  \end{figure}

The closure delays of triangles with all three baseline lengths\footnote{Total baseline length is used
throughout the paper as a proxy for projected baseline length.}
smaller than 7100 km were investigated. In Figure
\ref{small_net1}, the closure delays of these triangles are shown as
red circles connected by dashed lines. The time in hours on the X
axis is the GMST time of the observation
for each triangle epoch to show observations over the fifteen days in an 
overlapping 24-hour plot. The same technique is applied in the following figures of closure delays. 
There are in total 6611 closure delays for these
small triangles, and the standard deviation is about 21.7~ps.
Figure \ref{small_net2} geographically shows these small triangles\footnote{We should notice that there are more short baselines  
than shown in Figure \ref{small_net2}, since there are some
isolated short baselines unable to form such a small triangle.}. 
The magnitudes of the
closure delays of these triangles are mostly smaller than 40 ps. 
There are 24 closure delays with magnitudes
larger than 0.1 ns, and if we constrain the small
triangles to having baseline lengths shorter than 7000 km,
the magnitudes of the closure delay are all smaller than 0.1~ns.

\subsection{Triangles with one long and common baseline}

\begin{figure}[tb]
\begin{center}
 
\includegraphics[width=0.49\textwidth]{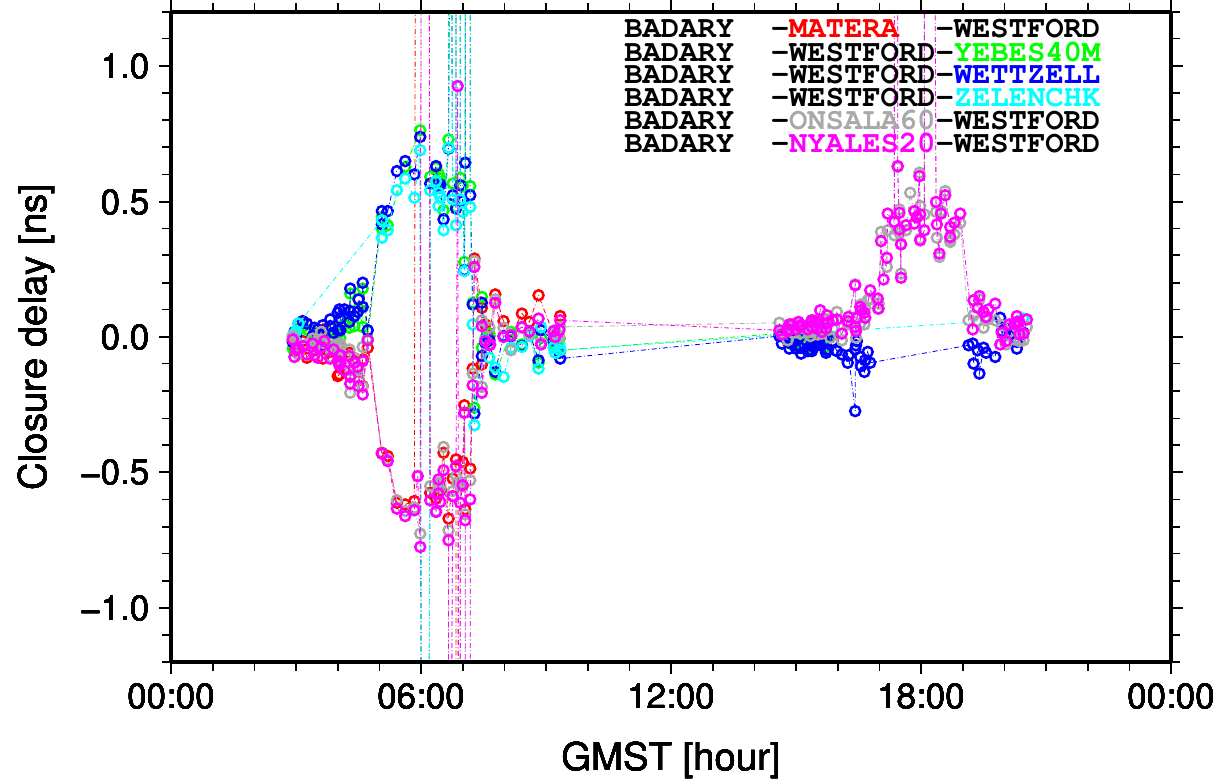}
  \caption{The closure delays of triangles with BADARY and WESTFORD and one of the six European stations.
  Each color corresponds to the closure delays of triangles of one kind of geometry, with the name shown on the upper right corner in the same color
  for the European station in that triangle. Points for the triangles with the same
  geometry are connected with dashed lines.}
  \label{s0642_bawe}

\end{center}

  \end{figure}

\begin{figure}[tb]
\begin{center}
 
  \includegraphics[width=0.48\textwidth]{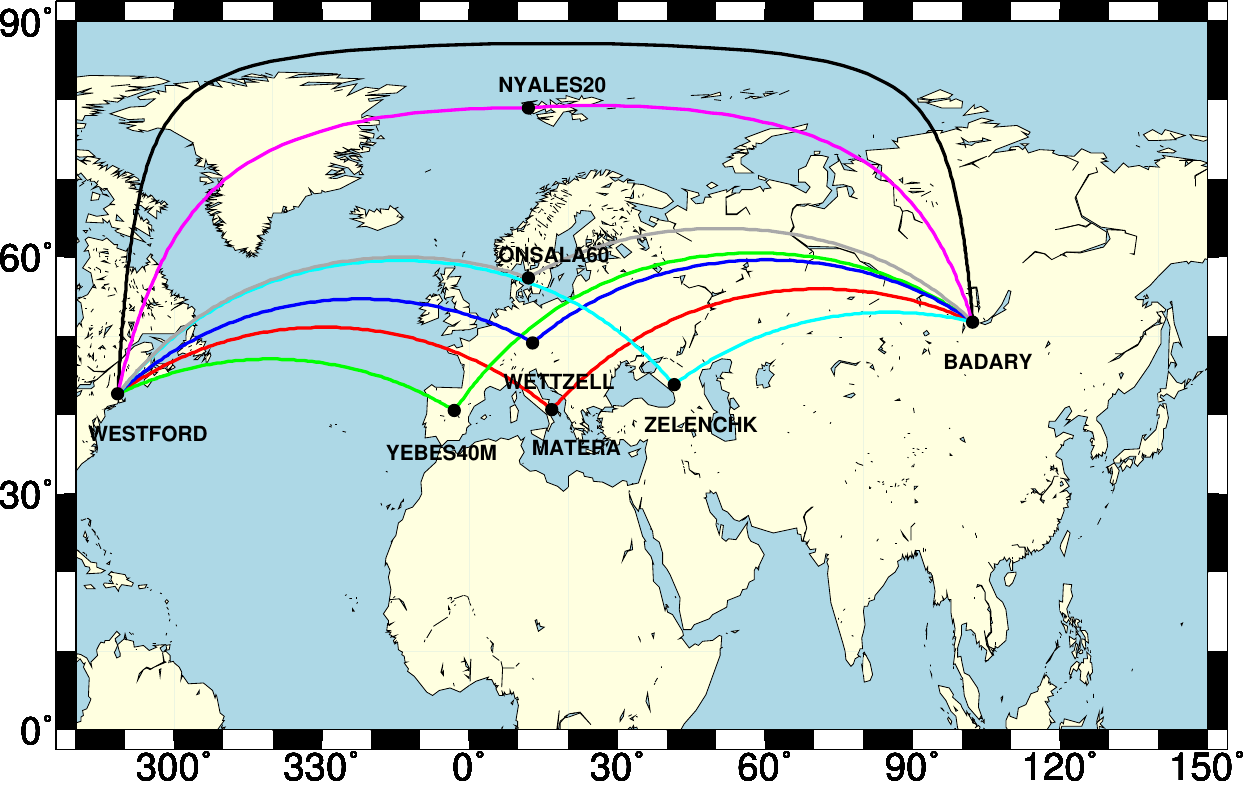}
  \caption{The geographic locations of BADARY, WESTFORD, and six central European stations. The black line
  is BADARY--WESTFORD; the six pairs of baselines connecting the stations BADARY and
  WESTFORD are shown with the same colors with that of the closure delays that they contribute to in Figure \ref{s0642_bawe}. }
  \label{s0642_bawe_net}

\end{center}

  \end{figure}

The closure delays of triangles that contain one long baseline and 
two short baselines were investigated. In the CONT14 network, six stations are densely
located in the European region so that one can make use of them to
construct a good set of triangles for this investigation. Figure
\ref{small_net2} suggests that the long baseline should be BADARY--WESTFORD, 
since the European region locates almost in the middle of
it. The length of baseline BADARY--WESTFORD is approximately 8700
km. ZELENCHK connects to BADARY with the baseline length of
approximately 4400 km, but to WESTFORD about 7800 km, a little longer than 7100 km; 
the other five European stations connect to both
with baselines shorter than 7100 km. 
Figure \ref{s0642_bawe} shows the closure delays with respect to the GMST time of the resulting six kinds of
geometry using six colors. Figure \ref{s0642_bawe_net}
shows the geographic positions of these eight stations and the
triangles with these six different geometries. 
Aside from the points outside the $Y$ axis limits, these
six sets of closure delays roughly have the same pattern, two peaks
happening around 03:00 and 15:00 with magnitude of approximately 0.5--0.7
ns. When WESTFORD is the third station in the triangle, the sign of
the peaks is opposite to that when WESTFORD is the second\footnote{By the naming
convention, in this case, WESTFORD can never be the first station in a baseline, and BADARY is always the first station.}.
Considering that baseline BADARY--WESTFORD is the only common one in these triangles and that triangles
with baselines shorter than 7100~km show no closure delays larger in magnitude than 0.2~ns,
it is reasonable and obvious to conclude that the systematic
variation in the closure delays of these six kinds of
geometry is predominately related to the baseline BADARY--WESTFORD. 
In addition, the closure delays of the nanosecond order and
larger, which are out of the axis limit in the plot, only happened
around the same GMST time of the peaks.

\subsection{Triangles with the shortest baseline}

\begin{figure}
\begin{center}
 
  \includegraphics[width=0.48\textwidth]{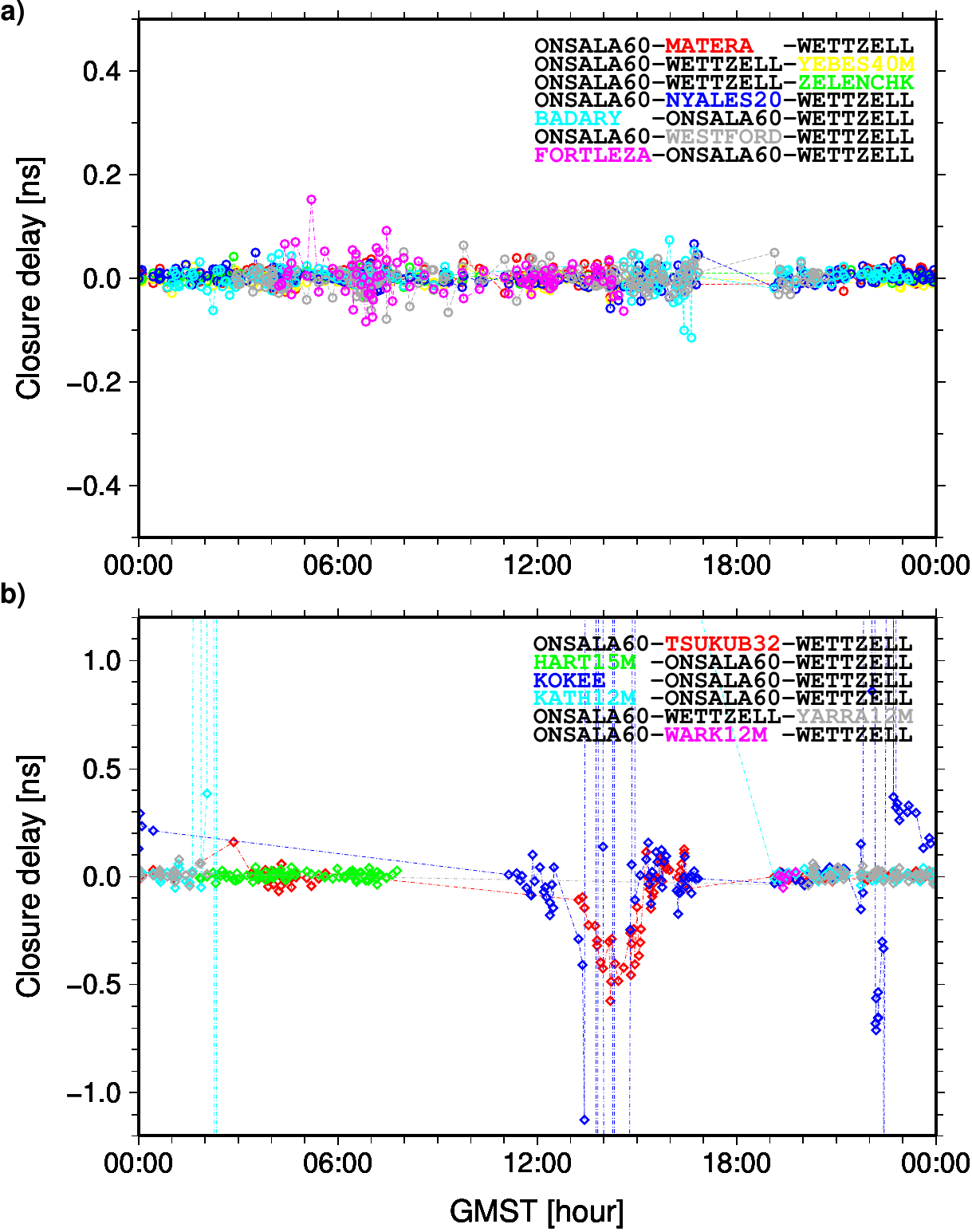}
  \caption{The closure delays of all triangles that contain the baseline ONSALA60--WETTZELL.
  Since 15 stations observed this source, there are 13 station available to join these two stations to construct a triangle.
  Subplot a shows the closure delays with seven different third station in different colors, while
  subplot b shows with a larger axis limit, closure delays from triangles using the remaining six stations. The names of 13 different kinds
  of geometry in both subplots are presented in the numerical order of their longest baseline lengths. For each geometry,
  the name of the third station is presented in the color of its closure delay. }
  \label{s0642_onwe_2}

\end{center}
  \end{figure}

As we know, the atmospheric and clock errors limit our ability to examine the structure effect on a single baseline on its own. 
The investigation in Section 3.2 above demonstrates the effect of source structure 
introduced by one long baseline by using two short baselines in the closure triangle
to cancel out the atmospheric errors, clock errors, and so on, in the long baseline.
The closure delays of all triangles including the smallest
baseline in the CONT14 network, ONSALA60--WETTZELL, were
investigated. They are shown in Figure \ref{s0642_onwe_2} by two
subplots with different scales. As there were 13 stations available
to form a triangle with baseline ONSALA60--WETTZELL, we have 13 different
kinds of geometry for this case. The closure delays were 
classified and sorted by the geometry of each triangle. In this figure, the names of 13
different geometries are presented in the numerical order of
the longest baseline length in the geometry, and the colors for the
names of the 13 stations correspond to those of the closure delay in
each of the subplots. Most closure delays of these triangles are very
small, at the level of tens of picoseconds, if the longest baseline in the
triangle is shorter than 7600 km. The magnitudes of the closure
delays become larger as the baseline lengths in the triangles continue
to increase, and eventually systematic variations appear. In this
case, the variations are much more complicated than that for
triangles with only one long baseline as shown in Figure
\ref{s0642_bawe}. Closure delays larger than 1 ns appear only when
there are long baselines in the triangle and where apparent systematic 
variations for short baseline triangles occurred. 

\section{Method to solve triangles}

As discussed in Section 2, the measured closure delay is due to the observational
noise and the source structure effect that is baseline
dependent. Then

    \begin{equation}
     \label{eq_closure3}
\tau_{abc}=\delta \tau_{ab}+\delta \tau_{bc}- \delta \tau_{ac} + \sum_{i=1}^{3} \epsilon_{i},
    \end{equation}
where $\delta \tau_{ab}$, $\delta \tau_{bc}$, and $\delta \tau_{ac}$
are the source structure effects on the three baselines, and
$\epsilon_{i}$ is the measurement noise in the $i$-th baseline
of that triangle. It is reasonable to assume here that
measurement noises are random at the level of a few picoseconds, which has been demonstrated 
by the closure delays of unresolved sources.

\begin{figure}[tb]
\begin{center}
 
  \includegraphics[width=0.46\textwidth]{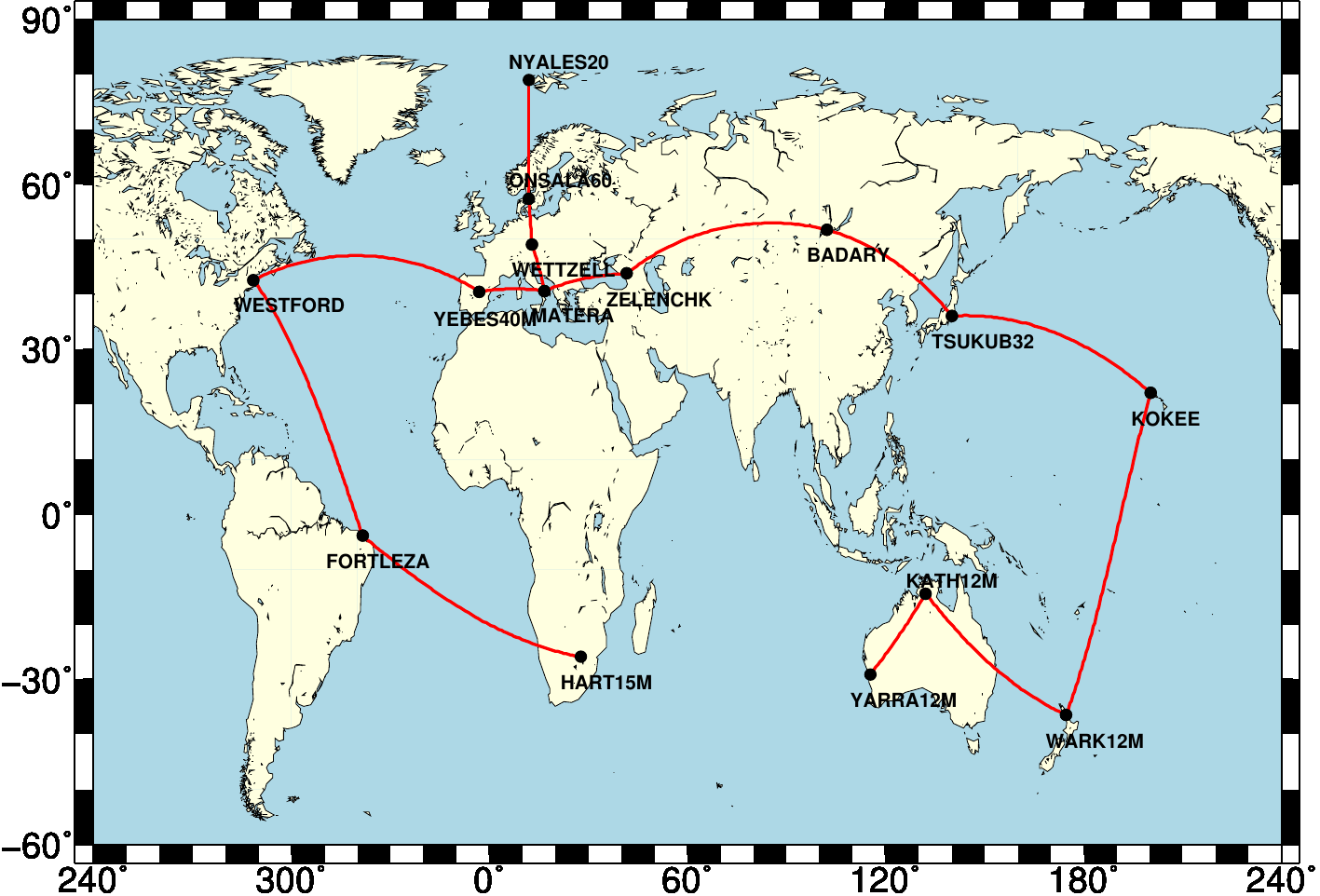}
  \caption{An example of a complete connection of the fifteen-station network by a minimal number
  of $<$ 7100 km long baselines. To solve for the baseline delays using closure delay in one scan, these selected
  baselines are set to have zero source structure effect in that scan.
  This kind of connection and its role is designated as connection in the paper.}
  \label{s0642_datum}

\end{center}

\end{figure}

In an array of $N$ stations, there are at most $N(N-1)/2$ baselines
and $N(N-1)(N-2)/6$ closure delay relations. But only $(N-1)(N-2)/2$ of these
relations are independent \citep{{pea84}}. There are therefore $(N-1)$ too few closure delays to 
determine the source structure effect for each standard baseline-delay observable, and 
an independent estimate of the structure delay on $(N-1)$ baselines has to be derived. 
The investigations in Section 3 give insights to this and 
the proposed method uses the assumption that source 0642+449 is
a point-like source with respect to the baselines shorter than a
certain value. We then choose 7100 km as the threshold and
select a necessary and minimum number of $<$ 7100 km
baselines to connect as many stations as possible in each scan.
The structure effects on these selected baselines are assumed to be zero. Taking
a fifteen-station array as an example, ideally there are
fourteen baselines shorter than 7100 km connecting fifteen
stations as a complete connection shown in Figure \ref{s0642_datum}. 
Setting the source structure effect on
these selected baselines to be zero allows us to solve for the structure effects
on other baselines utilizing the closure delays.

To find such a kind of a complete connection could be very challenging in some
cases. This is the case for the southern stations. Three southern
stations, KATH12M, YARRA12M, and WARK12M, can be connected to the
whole network through such a connection only if at least one of the
baselines KOKEE--WARK12M (6600 km) and KATH12M--TSUKUB32 
(5500 km) is available. If both of these two baselines get lost in a
scan, it is thus impossible 
to solve all the big triangles related to these three
stations. This is also one of the reasons for choosing 7100 km
as the threshold for the baseline length of the connection, considering
that the shortest baselines related to station HART15M, are
HART15M--MATERA (7033 km) and FORTLEZA--HART15M (7025 km).
Next step is to solve as many triangles as possible based on the
connection with the threshold of 7100 km scan by scan.

\section{Results of source structure effect}

\begin{figure}[tb]
\begin{center}
 
  \includegraphics[width=0.50\textwidth]{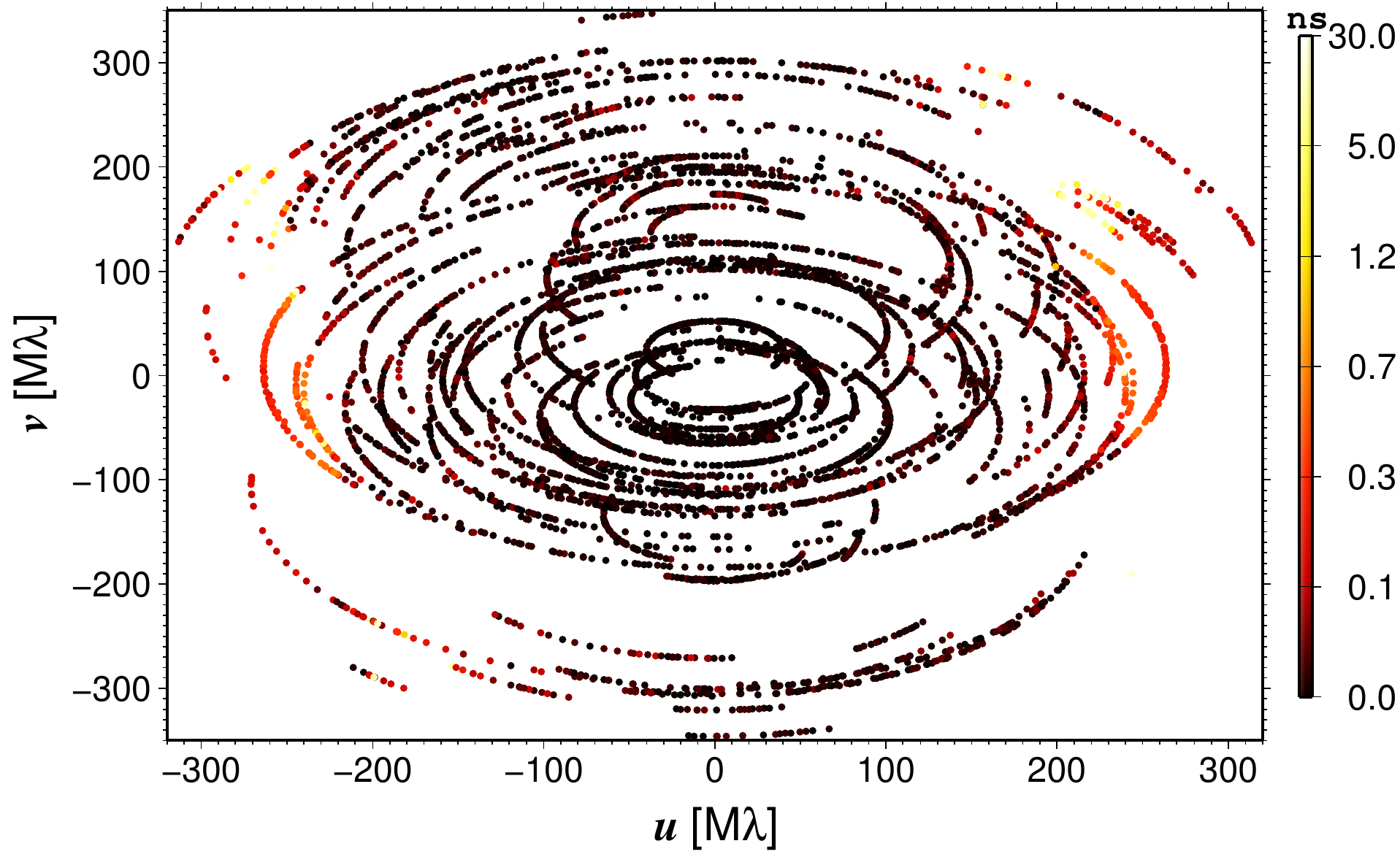}
  \caption{Coverage of the Fourier domain ($uv$ coverage) of the CONT14 observations of 0642+449 at
  8.4 GHz ($\lambda$ = 3.6 cm), plotted in units of Mega lambda. Color marks the absolute magnitude of
  the source structure effect on each of 8492 observables
  derived from closure delays by the method introduced in Section 4. }
  \label{uv0642_all}

\end{center}

\end{figure}

Finally, 16~941 triangles were solved and the source structure effects
on 8492 baseline delays were determined based on Equation
(\ref{eq_closure3}). In this solution, 2179
observables of short ($<$ 7100 km long) baselines were
assumed to have no source structure effect to build up the connection. Figure \ref{uv0642_all} shows the
$uv$ coverage with color marking the magnitude of the derived source
structure effect on each point. In general, the source structure has
a strong effect on long baselines at two opposite directions of approximately the $u$ axis. 
The results are divided into two groups for the
detailed study of learning the source structure effect.

\begin{figure}[tb]
\begin{center}
 
 \clearpage
 
\includegraphics[width=0.48\textwidth]{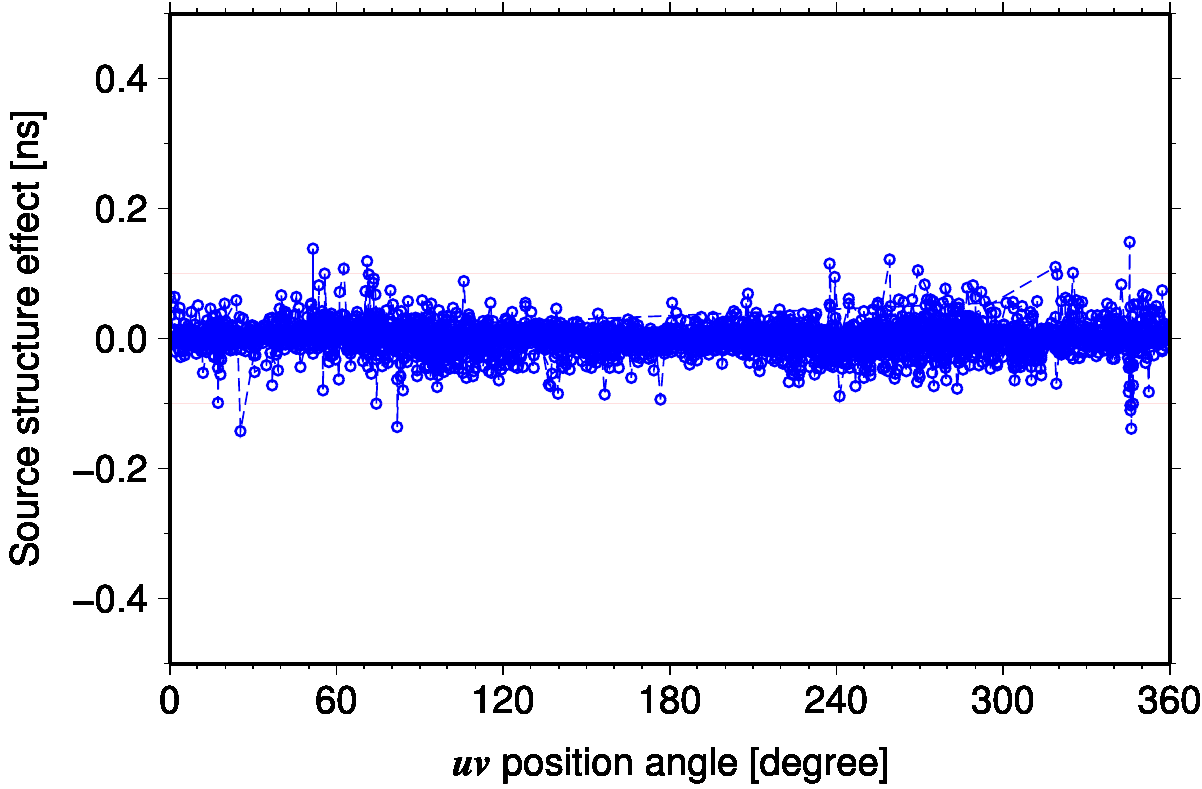}
  \caption{Derived source structure effects on 3443 observables of short baselines ($<$ 7100 km) from the
  closure delays, plotted as a function of the $uv$ position angle of baseline. }
  \label{s0642_s71}

\end{center}

  \end{figure}

First, apart from the observables selected to form the connection, the structure
effects on the remaining 3443 observables of short baselines were
estimated along with long baselines in the solution. The
estimated source structure effects on these baselines with respect to $uv$ position angle of baseline in the $uv$ plane 
(measured North-through-East) are shown in
Figure \ref{s0642_s71}. 
All these observables without exception have
very small source structure effects, and the standard deviation is
about 22.7 ps, which is close to that of the closure delays of
triangles with short baselines shown in Figure \ref{small_net1}. It
may be slightly larger than that of random measurement noises in
VLBI measurements, partly because the measurement noise in 2179
zeroed observables were propagated into the estimated ones.
There is no point larger than 0.2 ns. Therefore, the result
supports the assumption of no significant structure on short baselines 
that has been used for the connection.

\begin{figure*}
\begin{center}
 
               \includegraphics[width=0.45\textwidth]{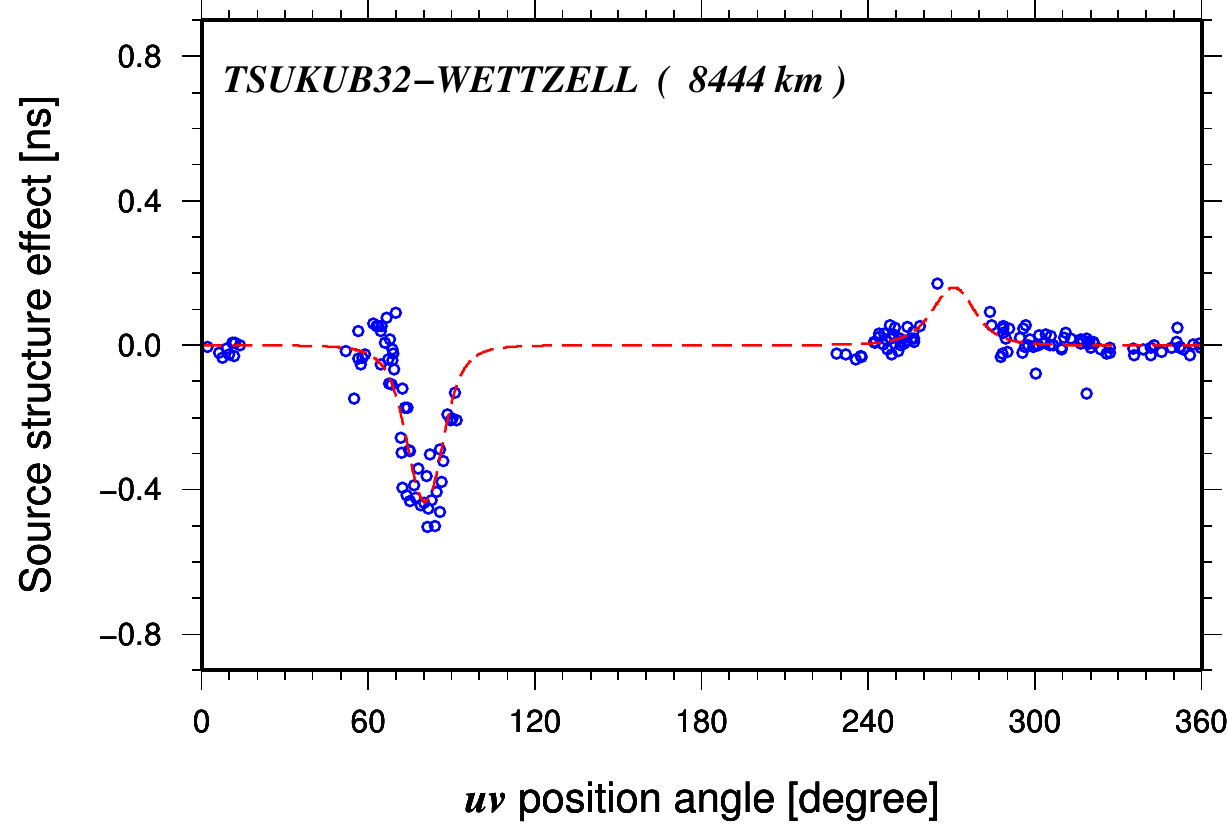}
         \includegraphics[width=0.45\textwidth]{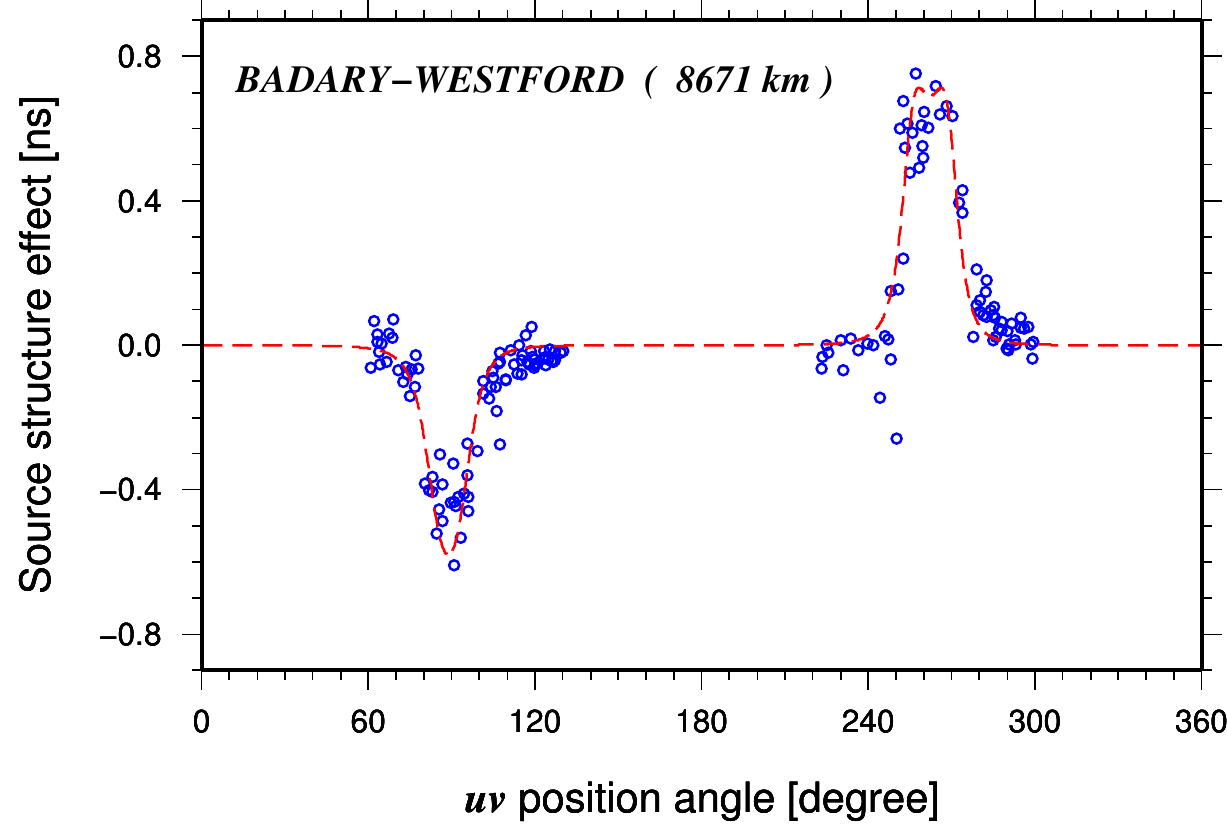}
           \includegraphics[width=0.45\textwidth]{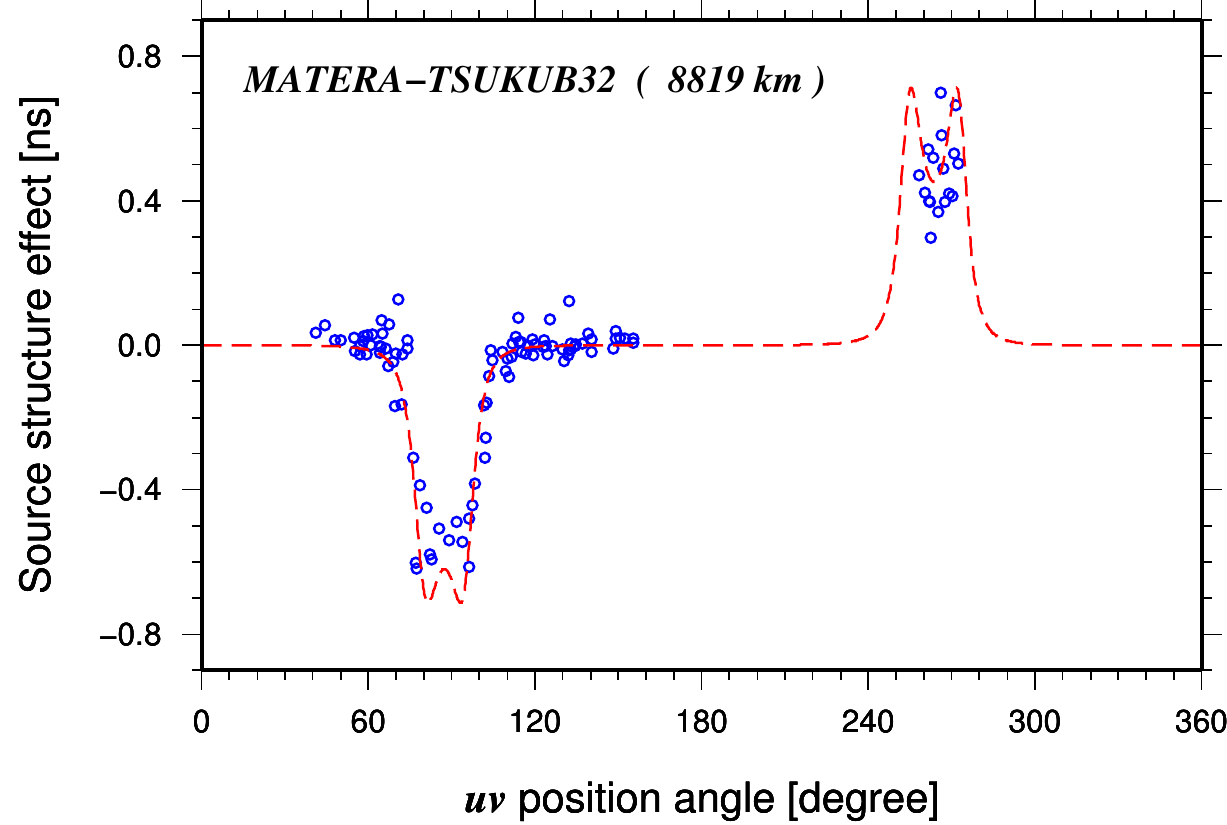}
             \includegraphics[width=0.45\textwidth]{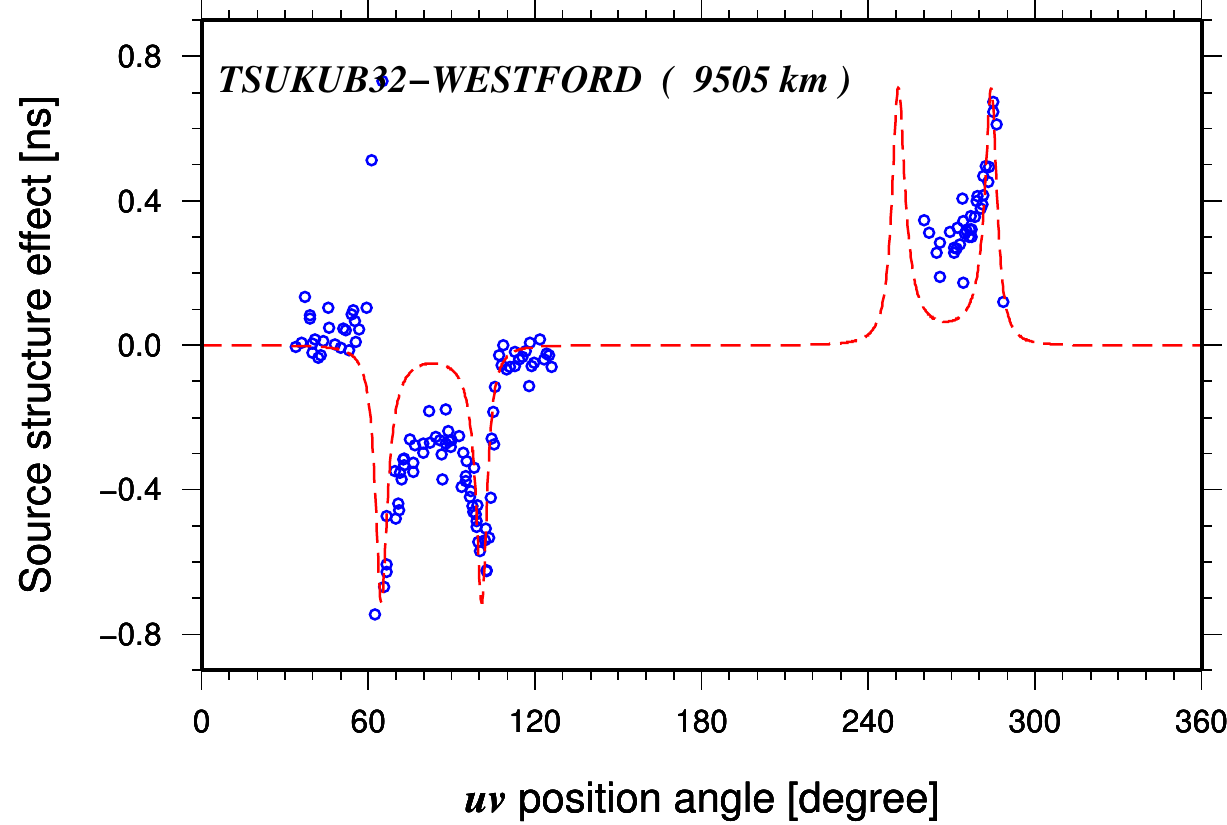}
         \includegraphics[width=0.45\textwidth]{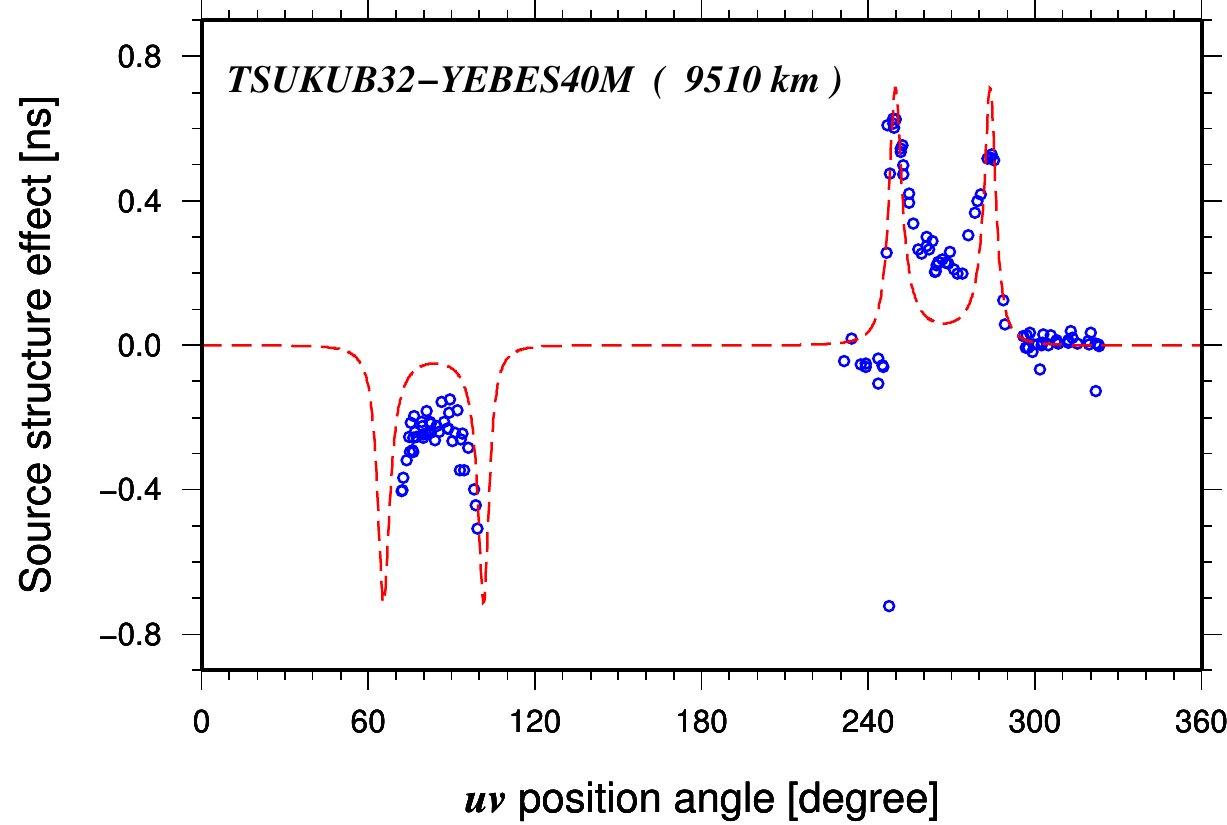}
           \includegraphics[width=0.45\textwidth]{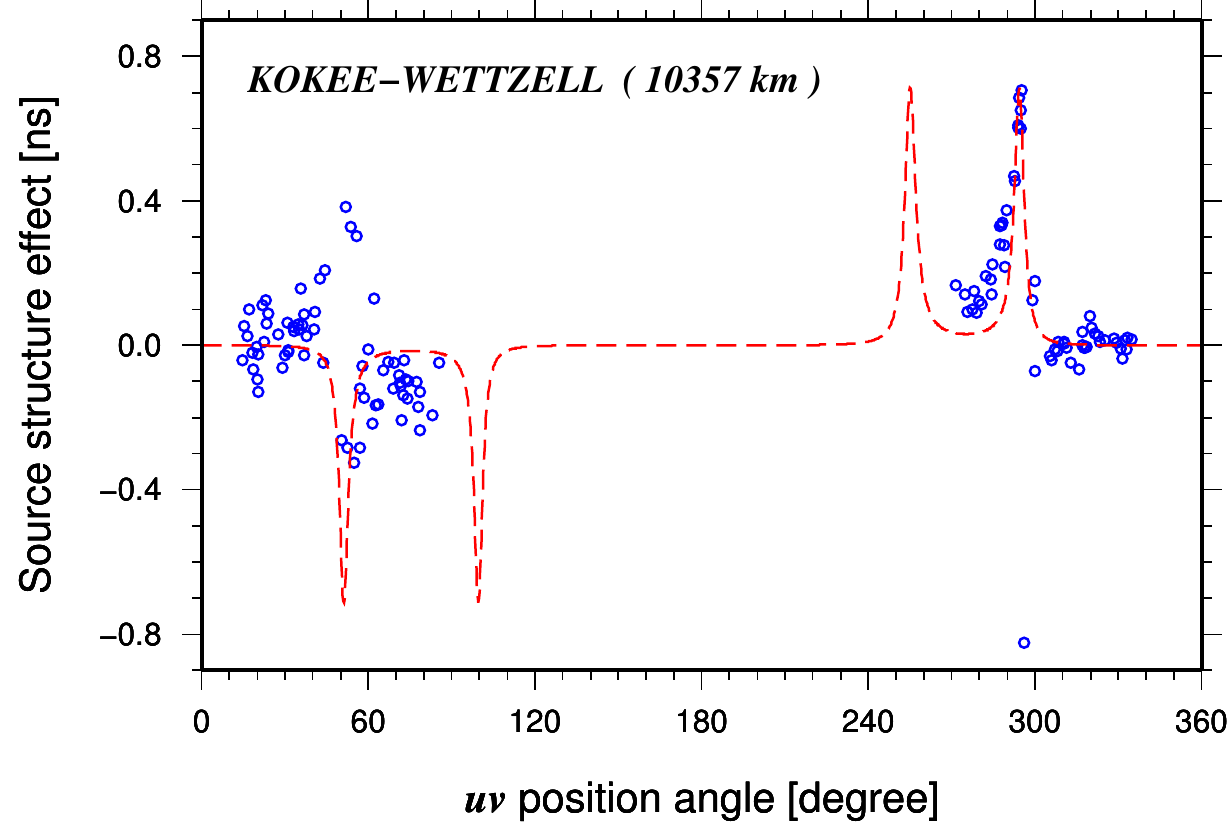}
             \includegraphics[width=0.45\textwidth]{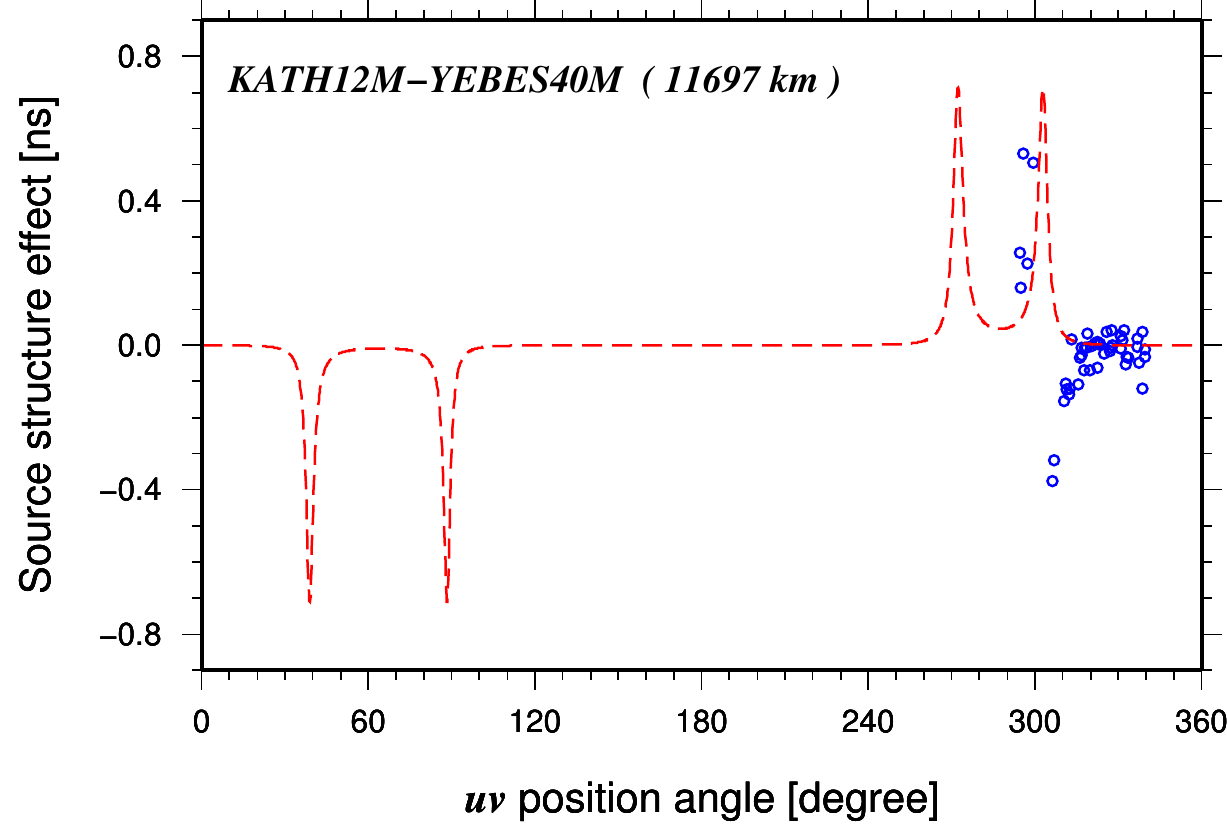}
               \includegraphics[width=0.45\textwidth]{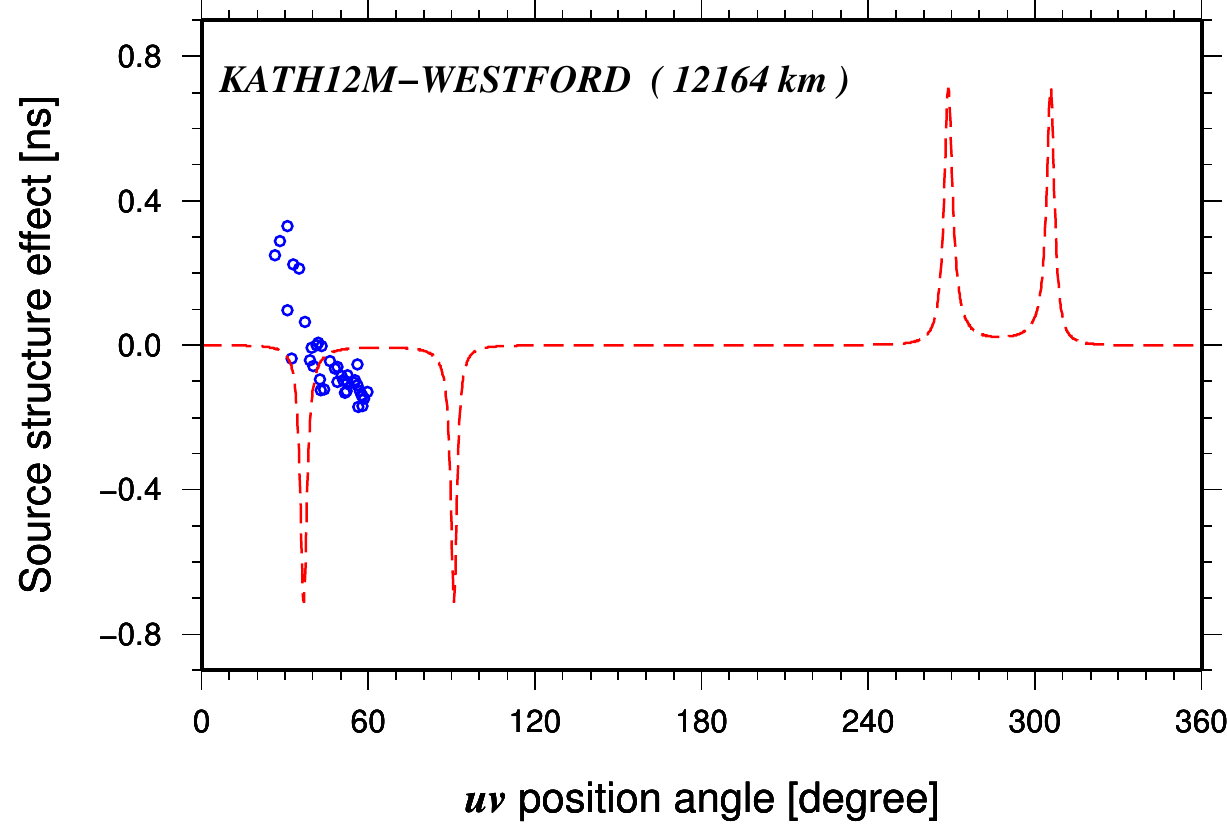}
  \caption{Derived source structure effects from the closure delay and the pattern of structure effects calculated from the two-component model as a function of the $uv$ position angle for eight baselines with significant variations. The baseline name along with the baseline length in brackets is shown for each of the plots. The blue dots correspond to the observation points, and the complete patterns of structure effects over one circle calculated from the model are shown in red dash lines.}
  \label{s0642_model}

\end{center}

  \end{figure*}

Secondly, the source structure effects for eight long baselines with
significant variations are presented one by one in Figure \ref{s0642_model} in blue dots. 
The patterns shown in these plots are incomplete over one circle of $uv$ position angle,
especially for the longest baselines, due to both the loss of common
visibility of the source and the vulnerability of losing link to
the connection. This indeed makes the identification of the source
structure extremely challenging. However, it is beneficial to
combine together the patterns in these plots. Two peaks for the
baselines with length from 8400 km to 9600 km can be
identified, appearing at the $uv$ position angles of approximately 90$\degr$
and 270$\degr$. The peak structure effect for these baselines is 0.7~ns.
Moreover, there are two characteristics of the source structure that can be identified.
First, apart from the points outside of the axis limit, the scatter of
the structure effect is rather small along the track of the 
variation curve, and this should imply that the source has multiple
point-like components rather than a flat brightness distribution
over the extended structure. Second, the variation is symmetric over
the $uv$ position angle of one circle, and the separation between the two
peaks is 180$\degr$. As a result, the source should have a symmetric
brightness distribution, for example, the multiple components are in
a straight line. Finally, points out of the axis limit happen exclusively
on baselines longer than 9000~km where significant source
structure effects occur.

%
%
%
%
%

\section{Modeling the source structure of 0642+449}                                 \label{sec:method}

The examples of a double-component source and a triple-component
source have been analyzed by \citet{tho80}, and a double-component
source has been studied in great detail by some simulations of
\citet{cha90b}. Based on these studies, we can determine that the
source exhibits a structure with two compact components on 
baselines with length of around 9000 km. The
ratio of the flux densities of the weaker component to
the stronger one, $K$, is about 0.92, since the peak structure effect
for these baselines is approximately 0.7 ns.

\citet{tho80} referred the source position in his model to the
middle point of the separation between the two components, while
\citet{cha90b} took the centroid point of brightness as the
reference point. The reference point does not matter for the closure
delay, but we  apply the centroid model here, because
we assume that observables of short baselines have zero source
structure effect, and in practice they are referred very closely to
the center of brightness. According to this model, the structure
delay $\tau_s$ on baseline $\mathbf{B}$ is given as

    \begin{equation}
     \label{eq_structure}
\tau_{s}=\frac{K(1-K)[1-cos(2 \pi R)]R}{f(1+K)[K^{2}+2Kcos(2\pi R)+1]},
    \end{equation}
where $f$ is the observing frequency, $R = \mathbf{B} \cdot \mathbf{S_{12}}/\lambda$, $\lambda$ is the wavelength, and
$\mathbf{S_{12}}$ is the relative position vector in the $uv$ plane of the weaker
component $P_2$ with respect to the stronger one $P_1$. This
expression only accounts for the change of the projected baseline on the $uv$ plane, i.e., the sky fringes.
It assumes that the flux densities of two components do not
change with frequency and observations are made on an infinitesimal bandwidth. 

Since the closure delay is not
sensitive to the reference point in the source structure
at all, this study lets all observables refer
to an unknown point, in this case to the centroid of brightness by the choice of the model, per scan. 
This reference point will finally be determined through VLBI data analysis.
We further assume that the source structure does not change within the fifteen days of observations.
A ``guess'' estimation for the unknown parameters in Equation (\ref{eq_structure}) was made from the results 
of baselines with lengths from 8400~km to 9600~km derived in Section 5. 
The final estimation was done 
by model-fitting all the closure delays with magnitude smaller than 1.0 ns 
based on the a priori values from the guess estimation.
The flux-density ratio is then estimated to be 0.916 $\pm$ 0.012, and the relative position
vector be $-$426 $\pm$ 12 $\mu$as in right ascension
and $-$66 $\pm$ 19 $\mu$as in declination. According to the
detected morphology, a baseline with length of 
7100~km has $R =$ 0.41 and the peak structure effect is only 13~ps, which explains the 
foundation of the assumption short baselines have no structure effect used
for the connection.

One can easily compute the source structure effect for each
observable based on Equation (\ref{eq_structure}) and the two-component model. 
The results for the eight baselines are shown in Figure
\ref{s0642_model} in red dash lines. By comparison, we found
that the model fits all the baselines with length smaller than 10~000~km, 
but it does not fit so well for some of the longest baselines, for
example baseline KATH12M--WESTFORD shown in Figure
\ref{s0642_model}, where the model and the estimated
structure effect just have different variations. After using the
model to correct the structure effect, the standard
deviation of the closure delay was decreased from 139 ps to
90 ps for all triangles with closure delay less than 1 ns, which
is a significant improvement.

\begin{figure}[tb]
\begin{center}
 
  \includegraphics[width=0.48\textwidth]{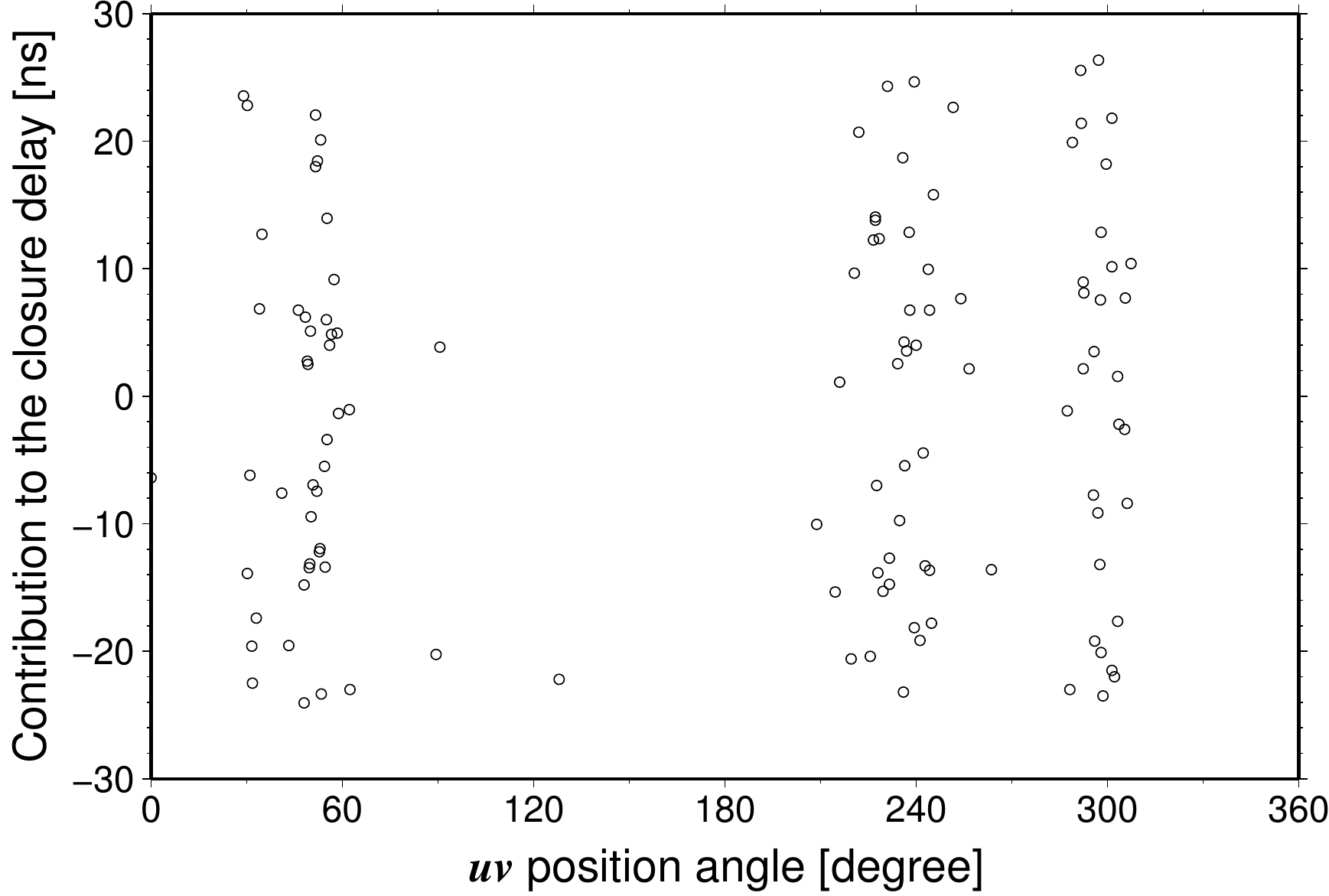}
  \caption{Distribution of the 117 observables with contribution to the closure delay larger than 1 ns. }
  \label{uv0642_angle}

\end{center}

  \end{figure}

\begin{table}
\begin{center}
\caption{Statistics of the 117 observables with contribution to the closure delay larger than 1 ns.} \label{table:1} \centering
\begin{tabular}{lrr}
\hline\hline    Baseline & Number & Length (km)\\
\hline
 FORTLEZA--KOKEE &   1 &  11063 \\
  HART15M--YARRA12M &   1 &   7848 \\
    KATH12M--NYALES20 &   1 &  10410 \\
    KOKEE--WESTFORD &   1 &   7676 \\
 WESTFORD--WARK12M &   1 &  11534 \\
 WESTFORD--YARRA12M &   1 &  12638 \\
 FORTLEZA--ZELENCHK &   3 &   8649 \\
   BADARY--FORTLEZA &   4 &  11154 \\
 FORTLEZA--TSUKUB32 &   4 &  12252 \\
  KATH12M--ONSALA60 &   4 &  10928 \\
  KATH12M--WETTZELL &   4 &  11026 \\
 TSUKUB32--WESTFORD &   4 &   9505 \\
   BADARY--WESTFORD &   5 &   8671 \\
  KATH12M--WESTFORD &   5 &  12164 \\
  KATH12M--YEBES40M &   5 &  11697 \\
    KOKEE--YEBES40M &   5 &  10687 \\
  KATH12M--MATERA &   6 &  10953 \\
    KOKEE--MATERA &   6 &  10894 \\
    KOKEE--WETTZELL &   9 &  10357 \\
    KOKEE--ONSALA60 &  15 &   9792 \\
    KOKEE--YARRA12M &  32 &   9498 \\
\hline\hline
\end{tabular}
\end{center}
\end{table}

Another solution was done to trace the observables that cause 749
triangles to have closure delays larger than 1 ns. It is much
easier to locate these observables actually, since the magnitude is
so large that all triangles related to them apparently have quite
large closure delays. The threshold of baseline length was then
reset to a little larger than the previous one, 8000 km for the
connection. The baseline with a length within this threshold should have a peak
structure effect of 0.12 ns, which will 
corrupt the determination of the source structure
effect for many observables, but facilitate locating the points
with large contribution to the closure delay. This solution allows
21~700 triangles to be solved, and 117 observables have been
identified with large contributions to the closure delay. Figure
\ref{uv0642_angle} shows the estimated closure delays of these 117
observables as a function of the $uv$ position angle. Table \ref{table:1}
shows the baselines lengths of these observables. All these points
systematically lie in three ranges of $uv$ position angle, 30$\degr$--60$\degr$, 220$\degr$--250$\degr$, 
and 280$\degr$--310$\degr$, and refer to baselines
with lengths larger than 7600 km, most of them larger than 9000~km. It leads us to conclude that these large closure delays are
more likely caused by the source structure as well. It is worth
noting that except for the baseline clock offset used to absorb
constant biases between baselines, no other parameters in VLBI data
analysis can absorb any part of the closure delay, which will finally impact the residuals. The
observables with a large contribution to the closure delay 
are always identified as
outliers during the data analysis. The study here shows that these
observables are not outliers but are systematically affected by source structure.

The limitations of the application of equation (\ref{eq_structure}) to the multiband 
group delay observable
need to be discussed here. Finite bandwidth for multiband group delay contributes to
errors in the modeled structure effects based on equation~(\ref{eq_structure})
because the structure changes with frequency and/or the
change in resolution $R$ as a function with frequency cause
$\tau_s$ to change with frequency. The X~band
observations in CONT14 use 8 channels spread over 720~MHz, for
a fractional bandwidth of 0.085.  We have performed numerical
simulations covering a large range of component separations,
baseline lengths and orientations, and K values to study this
effect.  For the case of $0642+449$, with $K \sim 0.9$ and the
separation about 0.5~mas, we find a median absolute deviation
between $\tau_s$ computed at the top and bottom of the
observed X~band frequencies, over the observed baseline
lengths and orientations, to be 2.6~fs, and the RMS deviation
to be 11~ps, for two components with the same spectral index.
With one component having a spectral index of 0 (typical for
an AGN core) and the other component having a spectral index
of $-0.7$ (typical for an optically thin jet component), the
median absolute deviation and RMS are 40~fs and 14~ps,
respectively.  Hence, the finite bandwidth of the CONT14
observations analyzed here does not present a significant
violation of the infinitesimal bandwidth assumption of
equation~(\ref{eq_structure}) for this source.  Furthermore, our simulations
show that having reasonable different spectral indices for the
two components contributes significantly less to differences
in the structure delay with frequency than the changes in
interferometer resolution ($R$) as a function of frequency
(14~ps for spectral index and $R$ changes together, compared
to 11~ps for $R$ changes alone).


\section{Conclusion and discussion}

Closure delay analysis has several advantages for geodetic VLBI. First, it
directly studies structure effects from the geodetic observables---the 
multiband group delays---themselves, the quantities actually
used to determine station position and motions, Earth
orientation parameters, and other astrometric/geometric parameters. 
Second, closure delay analysis can serve as
an indicator to evaluate the performance of any structure model,
whether that model is determined by fitting closure delays or through
an imaging analysis. Third, model fitting of closure delays can
identify the strongest components of a source while ignoring weak
components that do not significantly alter the group delay,
simplifying the analysis procedure. In contrast to many astrophysical
studies where the weak components are of interest, geodetic VLBI
analysis can often ignore such components. Fourth, a closure delay
analysis can save time and effort in both processing and software
development for geodetic purposes compared to standard VLBI imaging.

We showed that closure delays can determine the magnitude of the
measurement noise in the geodetic VLBI observables, and thereby also
determine the magnitude of structure effects in the geodetic VLBI
observables. We also showed that large closure delays, even closure delays
in excess of 10~ns, are related to source structure effects, and that
the underlying delay measurements are not caused by simple measurement
errors. We demonstrated for the first time that source structure can
be obtained from the closure \emph{delays} as opposed to closure
phases or closure amplitudes from visibility data. For sources that
are reasonably compact on short VLBI baselines, we can
simply and directly solve for (not fit) structure effects for the
entire VLBI network of baselines without any additional a priori
information. We also apply model fitting to determine source
structure, showing how closure delays can yield structure information
without the need for sources to be unresolved on short baselines. 
This method is relatively simple to implement
within existing geodetic analysis tools, uses input data
from the standard geodetic database files, and does not require
significant computational resources. For example, we can
compute structure models for all sources in the 15-day CONT14
campaign in a fraction of an hour, whereas our current imaging
analysis from the raw visibility data requires about 16~hours
to process one 24-hour segment of the CONT14 campaign on a
similar computer.

In an array of $N$ antennas, with $N(N-1)/2$ interferometer baselines, there are
$N(N-1)/2$ unknown structure effects to be determined and $(N-1)(N-2)/2$ 
independent closure delays as observables. Therefore, the fraction of the total
structure delay information available in the closure delays is
$(N-2)/N$. The ratio shows the benefits to be gained by increasing the number 
of antennas in the array; with only 4 antennas, 50\% of the structure delay information
is available, while for 15 antennas, as the case of CONT14 observations of source 0642+449,
the ratio increases to 87\%. From these observations, the source structure effect
is demonstrated at the level of each individual VLBI group delay for the first time.
The study reveals that at X band
(8.4 GHz) during the CONT14 sessions the source had two point-like components with a
flux-density ratio of 0.92, that is, almost equally bright. The position of the weaker component 
with respect to the stronger one is estimated to be $-$426 $\pm$ 12 $\mu$as in
right ascension and $-$66 $\pm$ 19 $\mu$as in declination. Finally,
the standard deviation of the corrected closure delay was reduced by
36$\%$. This structure model agrees with the estimated source
structure effect on baselines with lengths less than 10~000~km
very well, but does not fit some of the longest baselines. 
There are mainly four reasons for that: (1) source
structure effects on these longest baselines with $R \ga$ 0.7 are
much more sensitive to the relative position of the two components;
(2) such long baselines only have a few observations over one circle of $uv$ position angle, making it statistically
difficult to identify the variation caused by the source structure; (3) there can be structure at smaller scale that 
shows structure effect on longest baselines but none on baselines shorter than 10~000~km;
and (4) the model of Equation (\ref{eq_structure}) is 
the derivative of the structure phase with respect to the observing frequency, 
while the multiband group delay is derived from the linear estimation of the observed phase
over 8 channels spanning about 0.7 GHz---application of this model
formultiband group delays introduces errors in the structure effect, which may have larger impacts when
the baseline length is longer than 11~000~km with $R \ga$ 0.7. 
Due to this inadequateness of the model for multiband group delays, the flux-density ratio $K$ may have been 
underestimated.

In 1992 and 1995, this source was observed to have a compact ``core-jet"
morphology with the resolution of several milliarcseconds by \citet{gur92} and
\citet{xu95}, respectively. Recently, space VLBI (RadioAstron)
observations of this source at 1.6 GHz in 2013 with a resolution of
0.8 mas, $\sim$ 4 times better than that of ground VLBI images at
this wavelength, found that this source has two compact cores
separated by 0.76 mas with a position angle of 81$\degr$ in the sky plane \citep{lob15}.
Since space VLBI observations were made fourteen months earlier than CONT14 observations, one
may not expect they were observing the same blob, but the position angle of two components 
should be approximately in the same direction. 
Our result domnstrates that the two components are in the direction of position angle about
261.2$\degr$, which is the same direction detected by space VLBI.
The source 0642+449 did not exhibit a significant structure effect due to a frequency
dependence of the flux densities of the two components, which has a completely different
pattern, such as more peaks over 24 hours for baselines with $R \approx
0.5$. Our study shows a similar
structure of this source with a resolution comparable to that
of the space VLBI, demonstrating the feasibility of the application
of astrometric observables for the \deleted{direct} study of the source structure
with this method.

From the study by \citet{ber11}, we expect polarization 
leakage to affect the multiband group delay by less than 1.6 ps for
90~\% of the observations. General leakage of LCP into RCP for the 
geodetic observations will result in a baseline-dependent bias. For 
the LCP part of the Stokes I emission that leaks into RCP, these 
biases are constant in time and baseline orientation for a given 
station pair, and do not explain the large, systematic, and source-dependent effects. \citet{hom06} 
showed that at VLBI resolutions, the fractional circular polarization 
of AGN core and jet components is typically less than about 1~\%. 
Supposing that different baseline orientations constructively add/subtract 
the phases from two components that are circularly polarized at the 1~\% 
level, the change in delay caused by circularly polarized source 
structure would only be 2~\% of the change in delay for the Stokes 
I emission. We therefore expect that polarization effects are 
negligible for this study, although they may be important to reach picosecond accuracies.

The large closure delays have also been effectively traced, which
reveals that most observables erroneously identified as outliers in VLBI
data analysis are in fact exposed to the source structure effect and
this effect could be at the level of tens of ns in some occasions
even for a radio source with a rather flat spectral index. This
cannot be explained yet by the model. This
needs to be studied in the future to find the explanation from
astrophysics, while for astrometric VLBI we should schedule routine
observations more effectively to exclude this kind of radio source, 
or to observe it without such long baselines only if the two components
do not move with each other.

It is still challenging to implement the identified source structure
to correct the effect in VLBI data analysis. First, an accurate
model for multiband group delays to the level of at least 10 ps needs
to be derived. This model should be able to reduce the magnitudes of
the closure delays of triangles with the longest baselines to the
level of that of small triangles, a few tens of ps. Second, a careful re-study of the
linear combination of S and X band data with the presence of source
structure would be essential to have an accurate correction for the
source structure effect on the combined S-X observable. 
Moreover, astrometric/geodetic VLBI observables at S band are one order of 
magnitude noisier than that at X band,
which makes the source structure at S band almost unrecoverable, and
the structures at S and X bands are different. In general, structures 
at S band are much more resolved than that at X band.

How could our method be improved for the study of source structure? 
First, there should be more effective ways 
of deriving the structure effects on $(N-1)$ baselines.
\replaced{Without the limitation from the assumption used for the connection, 
our method by using the closure delay can be applied for 
the source consisting of compact components with any separation.
And this will allow us to directly correct the structure delay on each single delay observable
for geodetic VLBI data analysis.}{Due to the limitation from the assumption used for the connection, our method would confine to a small fraction of radio sources. But if one can develop a new way to break this limitation, our method may allow us to directly correct the structure delay on each single delay observable for geodetic VLBI data analysis.} 
Second, one might develop a new method of image reconstruction in an iterative way
other than modeling the structure delay. Then a non-linear estimation of 
the structure parameters from the closure delay 
has to be developed. The method can thus be extended to be used for
more general cases, complex or resolved sources. The rigorous method to correct the structure effect is to make images based on the standard VLBI imaging from the same observations and to correct the raw visibility phases for source structure in the geodetic VLBI analysis software prior to the multiband group delay fitting.\added{Even though this will need more work and resource compared to the current procedure of the routine VLBI data analysis, the geodetic VLBI should move onto it in the near future.} We are working on making images of CONT14 observations and the results will be presented in another paper.

This method could be of
great help to monitor the performance of radio sources for the historical VLBI observations and the VLBI
Global Observing System \citep[VGOS;][]{pet09}. In VGOS, there is a
global network of well-distributed stations and particularly
several twin telescopes. A wider range of baseline lengths from
hundreds of meters will be available, which then will allow
the source structure with a wide separation of compact components to be detectable. 
Moreover, if the point-like sources that are more likely to be observed in astrometric VLBI
begin to demonstrate structure, it is likely to roughly model their structures 
as consisting of compact components rather than a flat brightness
distribution. The source structure effect, as one of the main and
inevitable problems for the goals of the VGOS, can be expected to be
handled by this method to some extent. Besides the astrometric VLBI,
the method can provide benefits as imaging for the astrophysical
study of the source structure from continuous observations within
VGOS.

\acknowledgments

We thank the IVS for the CONT14 data used in this work, and David Gordon 
(Goddard Space Flight Center, USA) and Brian Corey (MIT Haystack Observatory, USA) for the helpful discussions 
of calculating closure delay from geodetic VLBI observations. This work was done while M.H.X.
worked as a guest scientist at GFZ, Germany, and supported by the National
Natural Science Foundation of China (Grant No. 11473057 and 11303077).

\clearpage


\begin{thebibliography}{}

\bibitem[Bertarini et al.(2011)]{ber11}
Bertarini, A., Roy, A. L., Corey, B., et al. 2011, Journal of Geodesy, 85, 715

\bibitem[Campbell, Schuh \& Zeppenfeld(1988)]{cam88} 
Campbell, J., Schuh, H., \& Zeppenfeld, G., 1988, 
in The Impact of VLBI on Astophysics and Geophysics, IAU Symposium No. 129, ed. J.M. Moran, M.J. Reid (Reidel: Dordrecht), 427


\bibitem[Charlot, Lestrade \& Boucher(1988)]{cha88} 
Charlot, P., Lestrade, J. F. \& Boucher, C., 1988, 
in The Impact of VLBI on Astophysics and Geophysics, IAU Symposium No. 129, ed. J.M. Moran, M.J. Reid (Reidel: Dordrecht), 33


\bibitem[Charlot(1990a)]{cha90a}
Charlot, P., 1990a, A\&A, 229, 51

\bibitem[Charlot(1990b)]{cha90b}
Charlot, P., 1990b, AJ, 99, 1309

\bibitem[Charlot(1993)]{cha93} Charlot P.,\ 1993, in EVGA 1993 General Meeting Proceedings, 171, ed. J. Campbell \& A. Nothnagel (Germany: Bad Neuenahr.)

\bibitem[Charlot(2010)]{cha10}
Charlot, P., 2010, AJ, 139, 1713


\bibitem[Feissel-Vernier(2003)]{fei03}
Feissel-Vernier, M., 2003, A\&A, 403, 105

\bibitem[Fey et al.(1996)]{fey96}
Fey, A. L., Clegg, A. W. \& Fomalont E. B., 1996, Astrophys. J., Suppl. Ser., 105, 299

\bibitem[Fey \& Charlot(1997)]{fey97} Fey, A.~L., \& Charlot, P.\ 1997, ApJ, 111, 95

\bibitem[Fey \& Charlot(2000)]{fey00} Fey, A.~L., \& Charlot, P.\ 2000, ApJ, 128, 17

\bibitem[Fey et al.(2015)]{fey15}
Fey, A., Gordon, D. ,Jacobs C. S., et al., 2015, AJ, 150, 58

\bibitem[Fomalont et al.(2011)]{fom11}
Fomalont, E., Johnston, K., Fey, A., et al., 2011, AJ, 141, 91

\bibitem[Gontier et al.(1993)]{gon93} Gontier A. M., Britzen S., Witzel A., et al.\ 1993, in EVGA 1993 General Meeting Proceedings, 167, ed. J. Campbell \& A. Nothnagel (Germany: Bad Neuenahr.)

\bibitem[Gurvits et al.(1992)]{gur92}
Gurvits, L.~I., Kardashev, N.~S., Popov, M.~V., et al.\ 1992, A\&A, 260, 82


\bibitem[Homan \& Lister(2006)]{hom06}
Homan, D. C., \& Lister, M. L. 2006, AJ, 131, 1262

\bibitem[Jennison(1958)]{jen58} Jennison, R.~C.\ 1958, 
MNRAS, 118, 276 

\bibitem[Kellermann \& Owen(1988)]{kel88}
Kellermann K. I. \& Owen F. N., 1988, 
Galactic and Extragalactic Radio Astronomy, (2nd ed.; The Observatory), Chapter 13

\bibitem[Lister et al.(2009)]{lis09} Lister, M.~L., Cohen,
M.~H., Homan, D.~C., et al.\ 2009, AJ, 138, 1874

\bibitem[Lister et al.(2013)]{lis13} Lister, M.~L., Aller,
M.~F., Aller, H.~D., et al.\ 2013, AJ, 146, 120

\bibitem[Lobanov et al.(2015)]{lob15}
Lobanov, A. P., G\'{o}mez, J. L., Bruni, G., et al.,
2015, A\&A, 583, A100


\bibitem[Ma et al.(1998)]{ma98}
Ma, C., Arias, E. F., Eubanks, T. M., et al., 1998, AJ, 116, 516


\bibitem[MacMillan(2007)]{mac07}
MacMillan, D. S. \& Ma, C., 2007, J. Geod., 81, 443

\bibitem[Malkin(2008)]{mal08}
Malkin, Z., 2008, J. Geod., 82, 325

\bibitem[Mo\'{o}r et al.(2011)]{moo11}
Mo\'{o}r A., Frey S., Lambert S., et al., 2011, AJ, 141, 178

\bibitem[Nothnagel(2015)]{not15}
Nothnagel, A., 2015, The IVS data input to ITRF2014, International VLBI Service for Geodesy and Astrometry, DOI: http://doi.org/10.5880/GFZ.1.1.2015.002

\bibitem[Ojha et al.(2004)]{ojh04}
Ojha, R., Fey, A. L., Johnston K. J., et al., 2004, AJ, 127, 3609

\bibitem[Ojha et al.(2005)]{ojh05}
Ojha, R., Fey, A. L., Charlot, P., et al., 2005, AJ, 130, 2529

\bibitem[Pearson 
\& Readhead(1981)]{pea81} Pearson, T.~J., \& Readhead, A.~C.~S.\ 1981, \apj, 248, 61 

\bibitem[Pearson 
\& Readhead(1984)]{pea84} Pearson, T.~J., \& Readhead, A.~C.~S.\ 1984, \araa, 22, 97 

\bibitem[Pearson 
\& Readhead(1988)]{pea88} Pearson, T.~J., \& Readhead, A.~C.~S.\ 1988, \apj, 328, 114 

\bibitem[Petrachenko et al.(2009)]{pet09} Petrachenko, B.,
Niell, A., Behrend, D., et al.\ 2009, NASA/TM-2009-214180, ftp://ivscc.gsfc.nasa.gov/pub/misc/V2C/TM-2009-214180.pdf

\bibitem[Petrov(2007)]{pet07} Petrov L.,\ 2007, in 18th EVGA General Meeting Proceedings, 141, ed. J. B\"{o}hm, A. Pany, H. Schuh (Germany: Geowissenschaftliche Mitteilungen.)

\bibitem[Piner(2007)]{pin07}
Piner, B. G., Mahmud, M., Fey, A. L., et al.,
2007, AJ, 133, 235

\bibitem[Plank et al.(2016)]{pla16}
Plank, L., Shabala, S. S., McCallum, J. N., et al., 2016,
MNRAS, 455, 343

\bibitem[Readhead 
\& Wilkinson(1978)]{rea78} Readhead, A.~C.~S., \& Wilkinson, P.~N.\ 1978, \apj, 223, 25 


\bibitem[Rogers et al.(1974)]{rog74} Rogers, A.~E.~E., 
Hinteregger, H.~F., Whitney, A.~R., et al.\ 1974, \apj, 193, 293 

\bibitem[Schuh \& Behrend(2012)]{sch12}
Schuh, H. \& Behrend D., 2012, J. Geodyn., 61, 68


\bibitem[Shabala et al.(2015)]{sha15} Shabala, S.~S.,
McCallum, J.~N., Plank, L., et al.,\ 2015, J. Geod., 89, 873

\bibitem[Sovers et al.(2002)]{sov02} Sovers, O.~J., Charlot,
P., Fey, A.~L., et al. \ 2002, in IVS 2002 General Meeting Proceedings, ed. N. R. Vandenberg \& K. D. Baver (Japan)


\bibitem[Tang \& R\"{o}nn\"{a}ng(1988)]{tan88} 
Tang \& R\"{o}nn\"{a}ng, 1988, 
in The Impact of VLBI on Astophysics and Geophysics, IAU Symposium No. 129, ed. J.M. Moran, M.J. Reid (Reidel: Dordrecht), 431

\bibitem[Taylor et al.(1994)]{tay94}
Taylor, G. B., Vermeulen, R. C., Pearson, T. J., et al., 1994, ApJS, 95, 345

\bibitem[Thomas(1980)]{tho80}
Thomas, J. B., 1980, ``An analysis of source structure effects in radio interferometry measurements", JPL publication 80-84, JPL, Pasadena, California Dec 15


\bibitem[Tornatore \& Charlot(2007)]{tor07}
Tornatore, V. \& Charlot P., 2007, J. Geod., 81, 469


\bibitem[Ulvestad(1988)]{ulv88} 
Ulvestad, J. L., 1988, 
in The Impact of VLBI on Astophysics and Geophysics, IAU Symposium No. 129, ed. J.M. Moran, M.J. Reid (Reidel: Dordrecht), 429

\bibitem[Whitney et al.(1971)]{whi71}
Whitney, A. R., Shapiro, I. I., Rogers, A. E. E., et al., 1971, Sci, 173, 225

\bibitem[Xu et al.(1995)]{xu95} Xu, W., Readhead, A.~C.~S.,
Pearson, T.~J., et al.,\ 1995, Astrophys. J., Suppl. Ser., 99, 297


\bibitem[Zeppenfeld(1991)]{zep91} Zeppenfeld G.,\ 1991, in EVGA 1991 General Meeting Proceedings, IV-16, (Netherland: Survey Department of Rijksw.)



\end{thebibliography}
\end{document}